\documentclass[pra]{revtex4}
\usepackage{amssymb}
\usepackage{graphicx}
\usepackage{epsfig,amsmath,graphics}
\usepackage{subfigure}

\renewcommand{\v}[1]{\ensuremath{\mathbf{#1}}}

\def\be{\begin{equation}}
\def\ee{\end{equation}}
\def\imagI{i}

\begin{document}

\title{An Atomic Gravitational Wave Interferometric Sensor in Low Earth Orbit (AGIS-LEO)}

\author{Jason M. Hogan}
\affiliation{Department of Physics, Stanford University, Stanford, California 94305, USA}

\author{David M. S. Johnson}
\affiliation{Department of Physics, Stanford University, Stanford, California 94305, USA}

\author{Susannah Dickerson}
\affiliation{Department of Physics, Stanford University, Stanford, California 94305, USA}

\author{Tim Kovachy}
\affiliation{Department of Physics, Stanford University, Stanford, California 94305, USA}

\author{Alex Sugarbaker}
\affiliation{Department of Physics, Stanford University, Stanford, California 94305, USA}

\author{Sheng-wey Chiow}
\affiliation{Department of Physics, Stanford University, Stanford, California 94305, USA}

\author{Peter W. Graham}
\affiliation{Department of Physics, Stanford University, Stanford, California 94305, USA}

\author{Mark A. Kasevich}
\email{kasevich@stanford.edu}
\affiliation{Department of Physics, Stanford University, Stanford, California 94305, USA}

\author{Babak Saif}
\affiliation{Space Telescope Science Institute, Baltimore, Maryland 21218, USA}

\author{Surjeet Rajendran}
\affiliation{Center for Theoretical Physics, Laboratory for Nuclear Science and Department of Physics, Massachusetts Institute of Technology, Cambridge, MA 02139, USA}
\affiliation{Physics Department, Johns Hopkins University, Baltimore, Maryland 21218, USA}

\author{Philippe Bouyer}
\affiliation{Laboratoire Charles Fabry de l'Institut d'Optique, Centre National de la Recherche Scientifique, Universit\'{e} Paris Sud 11, Institut d'Optique Graduate School, RD 128, 91127 Palaiseau Cedex, France}

\author{Bernard D. Seery}
\affiliation{NASA Goddard Space Flight Center, Greenbelt, Maryland 20771, USA}

\author{Lee Feinberg}
\affiliation{NASA Goddard Space Flight Center, Greenbelt, Maryland 20771, USA}

\author{Ritva Keski-Kuha}
\affiliation{NASA Goddard Space Flight Center, Greenbelt, Maryland 20771, USA}

\preprint{MIT-CTP 4175}

\date{\today}

\begin{abstract}
We propose an atom interferometer gravitational wave detector in low Earth orbit (AGIS-LEO).  Gravitational waves can be observed by comparing a pair of atom interferometers separated over a $\sim 30 \; \text{km}$ baseline.  In the proposed configuration, one or three of these interferometer pairs are simultaneously operated through the use of two or three satellites in formation flight.  The three satellite configuration allows for the increased suppression of multiple noise sources and for the detection of stochastic gravitational wave signals.  The mission will offer a strain sensitivity of $< \frac{10^{-18}}{\sqrt{\text{Hz}}}$ in the $50~\text{mHz}$ -- $10~\text{Hz}$ frequency range, providing access to a rich scientific region with substantial discovery potential.  This band is not currently addressed with the LIGO or LISA instruments.  We analyze systematic backgrounds that are relevant to the mission and discuss how they can be mitigated at the required levels.  Some of these effects do not appear to have been considered previously in the context of atom interferometry, and we therefore expect that our analysis will be broadly relevant to atom interferometric precision measurements.  Finally, we present a brief conceptual overview of shorter-baseline ($\lesssim 100 \; \text{m}$) atom interferometer configurations that could be deployed as proof-of-principle instruments on the International Space Station (AGIS-ISS) or an independent satellite.
\end{abstract}

\maketitle

\tableofcontents

\section{Introduction and Science Goals}
\label{Sec: IntroductionAndSources}

Gravitational waves (GWs) provide an unexplored window into the Universe.  All current astrophysical observations rely on electromagnetic waves.  Gravitational waves, by contrast, are sourced by energy, not charge, and can thus reveal entirely new types of sources that are difficult or impossible to observe electromagnetically, including black hole, white dwarf, and neutron star binaries.  Further, gravitational waves are not screened even by dense concentrations of matter and can thus be used to probe environments, including the very early universe, that are inaccessible to any other telescope.  Unlike electromagnetic waves, gravitational waves permit observations beyond the surface of last scattering.  Such observations could change our understanding of the fundamental laws of physics, shedding light on the birth of our universe in reheating after inflation or revealing evidence that our universe underwent a phase transition at extremely high temperatures.  Historically, the ability to observe new sections of the electromagnetic spectrum has led to many new discoveries.  Similarly, the gravitational spectrum seems likely to be filled with sources that will provide unique information about astrophysics, cosmology, and high energy physics \cite{Schutz:1999xj, Cutler:2002me, Maggiore:1999vm}.  However, it may be the unpredicted sources of gravitational radiation that have the greatest effect on our understanding of the universe.

We propose a low Earth orbit gravitational wave observatory based on atom interferometry, AGIS-LEO, which will be sensitive to signals in the frequency band $50~\text{mHz}$ - $10~\text{Hz}$, complementary to the reach of the LIGO and LISA instruments. The frequency band probed by AGIS-LEO is home to a variety of interesting astrophysical phenomena. Textbook sources of gravitational radiation are compact object binaries, which give a coherent, oscillatory gravitational wave signal. These binaries consist of systems where both components are compact objects such as white dwarfs, neutron stars, or black holes \cite{Cutler:2002me}.  They strongly radiate gravitational waves because they contain large-mass stars orbiting in close proximity, and the frequency of the emitted waves traverses through the AGIS-LEO frequency band as the binaries merge.

The lifetime of such a source is a strong function of the gravitational wave frequency and ranges from several thousand years for a source at 10 mHz to about a week for a source at 1 Hz. These observation times are significantly longer than the lifetime of a comparable source in the frequency band of LIGO. This long lifetime enhances the number of such sources in this frequency band relative to that of LIGO and increases their detectability owing to the increased observation time. Additionally, due to their large physical size, the mergers of white dwarfs end before they enter the frequency band accessible to LIGO but lie well within the frequency band of AGIS-LEO. Numerous such binaries are expected to lie in the Milky Way, and AGIS-LEO may be able to detect the brightest of these sources, such as a compact solar mass binary within $\sim 100$ kiloparsec.

Other major sources of gravitational radiation in the AGIS-LEO band include the inspirals of solar mass black holes into intermediate mass ($10^2 M_\odot \lessapprox M \lessapprox 10^4 M_\odot$) black holes out to distances of $\sim 1$ megaparsec and into massive ($10^4 M_\odot \lessapprox M \lessapprox 10^7 M_\odot$) black holes out to distances of $\sim 10$ megaparsecs. There are significant uncertainties in the expected merger rate of such objects, particularly because their detection is a challenge through electromagnetic methods. In fact, the detection of the gravitational waves emitted by these objects is one of the cleanest ways to observe them, and hence there is great scientific importance in achieving this goal.

The detection and parameter estimation of these mergers could play an important role in enhancing our understanding of the growth of structure in our Universe. For example, they can help uncover the nature of seed black holes that lead to the formation of massive black holes at the centers of galaxies.  These binary mergers will also test general relativity in the strong field regime - a regime which is currently inaccessible to terrestrial and solar system tests of gravity.  Furthermore, the detection of a large number of such binary mergers may permit their use as standard gravitational wave sirens (analogous to an electromagnetic standard candle), thereby providing a new, precision cosmological probe \cite{Schutz:1986gp}.

There are also several possible cosmological sources of gravitational radiation, described below, which are more easily observable in the low frequency band below 10 Hz.  Although these sources are by no means certain to exist, the discovery of any one would be incredibly important for cosmology and high energy physics.

We believe our universe began in a period of inflation; the subsequent reheating produced all the matter we see today.  This period of reheating after inflation produces gravitational waves which are potentially strong enough to be detectable \cite{GarciaBellido:2007af, Easther:2006vd, Dufaux:2007pt}.  The gravitational wave spectrum produced by reheating is expected to be peaked; the frequency of the peak is model-dependent and varies with the scale of reheating.  After redshifting, this peak can lie anywhere between roughly $10\text{ mHz}$ and $10^{9} \text{ Hz}$, so it may be possible for AGIS-LEO to observe gravitational waves from reheating.

A first-order phase transition in the early universe can produce gravitational waves through bubble nucleation and turbulence \cite{Kamionkowski:1993fg, Caprini:2007xq}.  A phase transition at the well-motivated TeV scale is likely to produce gravitational waves with a frequency today in a range near $100 \; \text{mHz}$.  In some models with new physics at the weak scale \cite{Grojean:2006bp}, including some supersymmetric \cite{Apreda:2001us} and warped extra-dimensional \cite{Randall:2006py} models, the electroweak phase transition can produce gravitational waves well above the threshold for detection by atom interferometers.

A network of cosmic strings produces a stochastic background of gravitational waves from cusps, kinks, and vibrations of the strings.  If such a network exists, calculations \cite{Polchinski:2007qc, DePies:2007bm} indicate that the gravitational waves produced may well be strong enough to be observable in the low frequency band from $\sim 100 \text{ mHz}$ to 10 Hz for cosmic string tensions as low as $G \mu \sim 10^{-8}$ (or lower for a more advanced detector that could be realized in the future).

There are many other possible sources of gravitational waves from fundamental physics in the early universe, including Goldstone modes of scalar fields \cite{Hogan:1998dv}, or radion modes and fluctuations of our brane in an extra dimensional scenario \cite{Hogan:2000is, Hogan:2000aa}. Pre-big bang \cite{Gasperini:2002bn} or extended \cite{Turner:1990rc} inflation could also lead to a strong gravitational wave signal in the relevant frequency range. Other astrophysical sources may lead to an interesting stochastic gravitational wave background (for a review see \cite{Maggiore:1999vm}).

\section{Mission Overview}

We propose to search for gravitational waves using light-pulse atom interferometry \cite{ref:Berman, varenna}.  (An alternative approach using confined lattice-hold interferometry will also be explored in Sec. \ref{sec:latHoldPhase}.)  In this section, we describe atom interferometry and how it can be used to make a sensitive space-based gravitational wave detector, AGIS-LEO.  We will also introduce the multiple-satellite configurations that arise naturally from the application of atom interferometry to the detection of gravity waves.

\subsection{Atom Interferometry}
\label{Sec: AI}
In a light-pulse atom interferometer (AI), an atom is forced to follow a superposition of two spatially separated free-fall paths.  This is accomplished by coherently splitting the atom wavefunction with a pulse of light that transfers momentum $\hbar \v{k_\text{eff}}$ to a part of the atom \cite{ref:Berman}, where $\v{k_\text{eff}}$ is the effective wavevector of the light.  As a result of the momentum recoil from the interaction with the light, the two halves of the atom wavefunction spatially separate along the light propagation direction, each following its own classical path due to the velocity difference.  Note that the interferometer paths considered here are all nominally one-dimensional along the direction of $\v{k_\text{eff}}$; unwanted displacements perpendicular to $\v{k_\text{eff}}$ arise only because of Coriolis forces present in low Earth orbit and will be discussed below (see Figs. \ref{Fig:FourPulseSequence} and \ref{Fig:FivePulseSequence} for examples).  When the atom wavefunction is later recombined, the resulting interference pattern depends on the relative phase accumulated along the two paths.  This phase shift results from both the free-fall evolution of the quantum state along each path and from the local phase of the laser, which is imprinted on the atom during each of the light pulses \cite{varenna}.  Since it precisely compares the motion of the atom to the reference frame defined by the laser phase fronts, the phase shift is sensitive to inertial forces present during the interferometer sequence.  To be concrete, the lasers are ultimately anchored to the frame of the AI sensor, so the AI phase shift is a record of the motion of the inertial atom with respect to the sensor frame as measured by the laser phase fronts, which act as a sub-micron periodicity ruler.  Equivalently, the atom interferometer phase shift can be viewed as a clock comparison between the time kept by the laser's phase evolution and the atom's own internal clock.  Sensitivity to gravitational waves may be understood as arising from this time comparison, since the presence of space-time strain changes the light travel time between the atom and the laser \cite{shortAGIS}.

Gravitational waves can be detected by recording the phase shift they induce in the atom interferometer.  Since this phase shift is also directly sensitive to unavoidable phase noise and vibration noise present on the laser light pulses that act as the atom beamsplitters and mirrors, a practical gravitational wave detector must be differential.  The concept of the AGIS-LEO instrument is to compare the phase shifts of two separate atom interferometers that are manipulated by the same laser, and that are therefore subject to the same laser noise.  The differential phase shift is still sensitive to the gravitational wave while the laser noise is suppressed as a common mode.

\begin{figure}
\includegraphics[width=7.0 in]{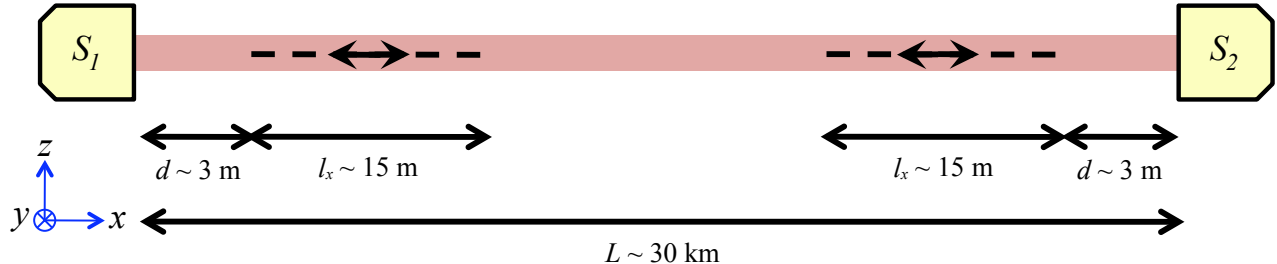}
\caption{ \label{Fig:satelliteSchematic} Schematic of one beam line of the proposed setup. Two satellites $S_1$ and $S_2$ house the lasers and atom sources. The dashed lines represent the path of length $l_x$ traveled by the atoms during the interferometer sequence. At their closest approach, the atoms are a distance $d$ from the satellites. The red band represents the paths of the lasers along the axis between the satellites.}
\end{figure}

\begin{figure}
\includegraphics[width=7.0 in]{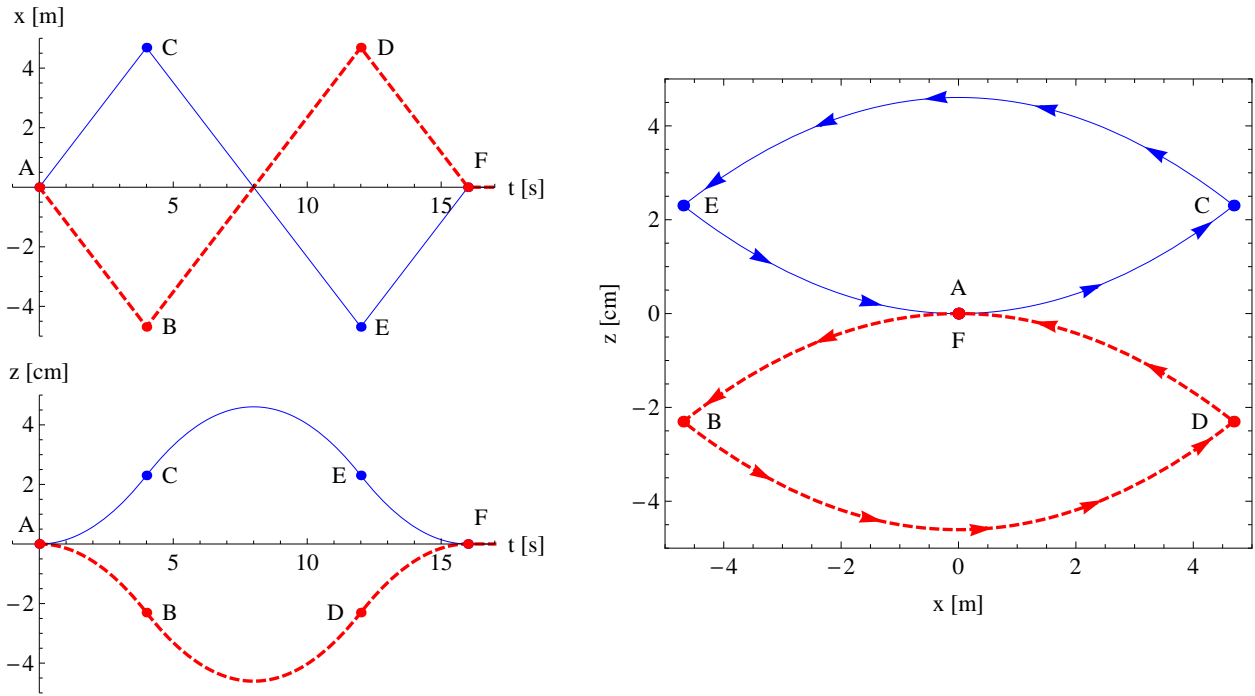}
\caption{ \label{Fig:FourPulseSequence} Four-pulse atom interferometer sequence using double-diffraction LMT beamsplitters under the influence of a leader-follower low Earth orbit rotation bias.  The upper arm of the interferometer is the solid (blue) path and the lower arm the dashed (red) path.  Interactions with the light pulses are shown as points labeled by letters A through F.  The interferometer laser pulses are directed along the $x$-direction with $\v{k}_i=\kappa_i k_\text{eff}\v{\hat{x}}$ for $1\leq i\leq 4$.  The normalized wavevectors have values $\kappa_1=1$, $\kappa_2=2$, $\kappa_3=2$ and $\kappa_4=1$.  The pulses occur at times $t_1=0$, $t_2=T$, $t_3=3T$ and $t_4=4T$ with interrogation time $T=4~\text{s}$.  See Fig. \ref{Fig:satelliteSchematic} for the definition of coordinate axes.}
\end{figure}

\begin{figure}
\includegraphics[width=7.0 in]{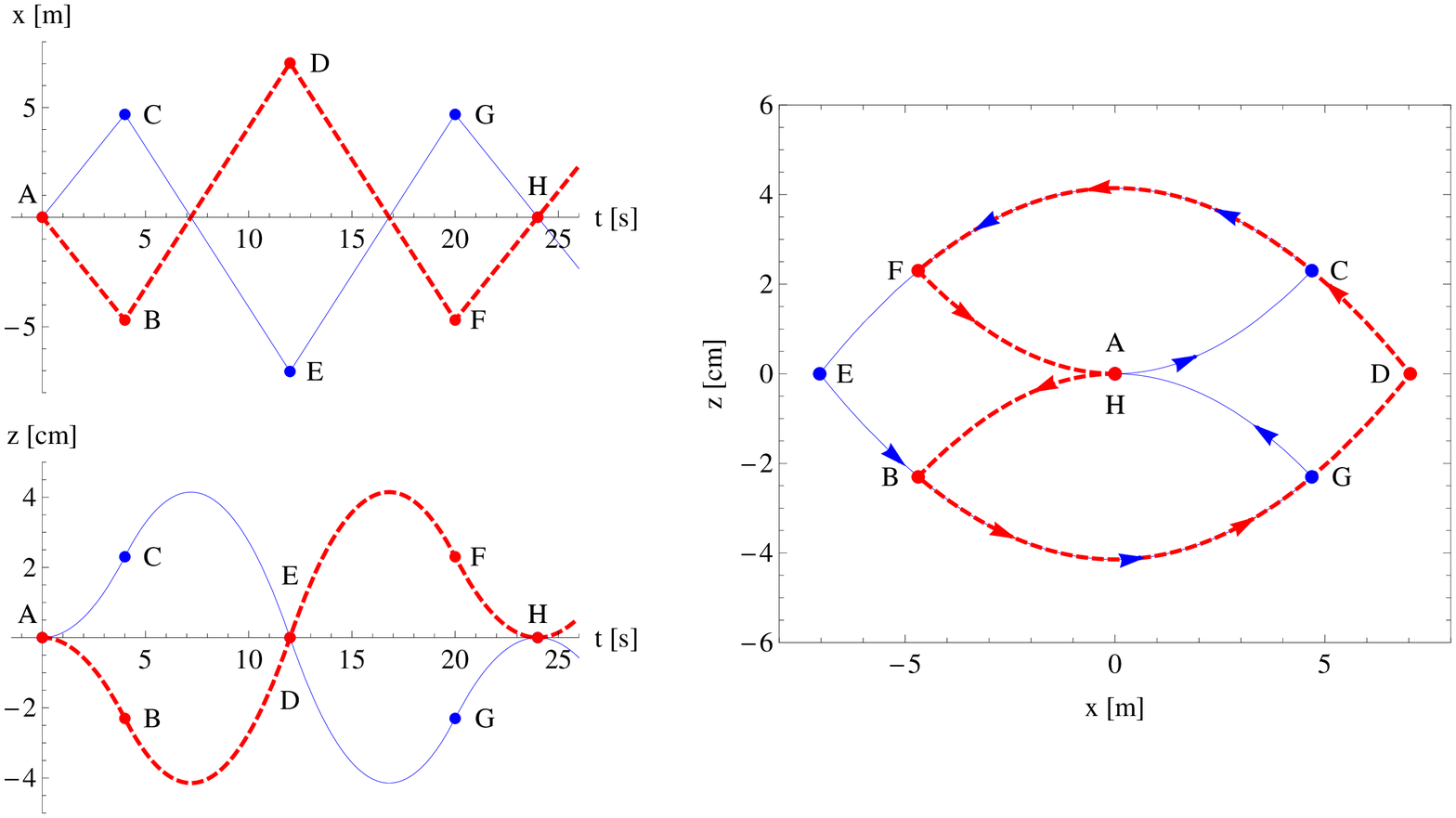}
\caption{ \label{Fig:FivePulseSequence} Five-pulse atom interferometer sequence using double-diffraction LMT beamsplitters under the influence of a leader-follower low Earth orbit rotation bias.  The upper arm of the interferometer is the solid (blue) path and the lower arm is the dashed (red) path.  Interactions with the light pulses are shown as points labeled by letters A through H.  The interferometer laser pulses are directed along the $x$-direction with $\v{k}_i=\kappa_i k_\text{eff}\v{\hat{x}}$ for $1\leq i\leq 5$.  The normalized wavevectors have values $\kappa_1=1$, $\kappa_2=\tfrac{9}{4}$, $\kappa_3=\tfrac{5}{2}$, $\kappa_4=\tfrac{9}{4}$ and $\kappa_5=2$.  The pulses occur at times $t_1=0$, $t_2=T$, $t_3=3T$, $t_4=5T$ and $t_5=6T$ with interrogation time $T=4~\text{s}$.  See Fig. \ref{Fig:satelliteSchematic} for the definition of coordinate axes.}
\end{figure}

Figure \ref{Fig:satelliteSchematic} illustrates the AGIS-LEO differential interferometer concept.  Two satellites $S_1$ and $S_2$ separated by baseline $L$ provide a pair of counter-propagating laser beams required to implement the necessary two-photon transitions commonly used for atom optics in alkali atoms \cite{kasevichChuAI, ref:giltner, ref:rasel}.  The atom interferometer regions (dashed lines) are each near one of the laser sources and have size $l_{x}$ along the laser direction (the $x$-direction).  As mentioned above, the atom wavepackets are separated along the laser propagation direction so that the paths of the interferometer arms are nominally one dimensional along $x$.  Both satellites are in orbit around the Earth, and for simplicity we take the orbital angular velocity vector to be $\v{\Omega}=\Omega_\text{or} \v{\hat{y}}$ (for a leader-follower orbit, as seen in Fig. \ref{Fig:greatCircles}).

The specific sequence of light pulses used to implement the atom interferometer is an important consideration since it affects both the gravitational wave response and the sensitivity to noise.  The standard three-pulse accelerometer pulse sequence ($\pi/2-\pi-\pi/2$) consists of an initial beamsplitter pulse, a mirror pulse a time $T$ later, and a final interfering beamsplitter pulse after an additional time $T$.  Such an interferometer is sensitive to gravitational waves as well as uniform acceleration; this interferometer geometry is discussed in detail in \cite{AGIS}.  However, in a leader-follower low Earth orbit (see Fig. \ref{Fig:greatCircles}) the three-pulse sequence cannot be used by the AGIS-LEO instrument because the large rotation bias prevents the overlap of the interferometer arms.  In order to observe interference and thereby measure a phase shift, the arms of the atom interferometer must overlap at the end of the pulse sequence within the size of the atomic wavepacket (i.e., within the spatial coherence length of the atom).  For $^{87}$Rb atoms, the coherence length of the atom's external degrees of freedom is $\Delta x\simeq \hbar/\Delta p \sim 10~\text{$\mu$m} \sqrt{\frac{100~\text{pK}}{\tau}}$ for atom ensemble temperature $\tau$.  The large Coriolis forces experienced in low Earth orbit can cause deviations to the nominal paths of the atom interferometer's arms that exceed this coherence length, causing the interferometer not to close.  To address this problem, AGIS-LEO uses multiple-pulse sequences with appropriate symmetry to facilitate interferometer closure.

An example of a pulse sequence that closes even in the presence of large Coriolis forces is the symmetric four-pulse sequence ($\pi/2-\pi-\pi-\pi/2$) shown in Fig. \ref{Fig:FourPulseSequence}, where the four laser pulses propagate along the $x$ direction with effective wavevectors $\v{k}_i=k_i \v{\hat{x}}$ for $1\leq i\leq 4$.  This interferometer uses two mirror pulses, resulting in a two-loop interferometer geometry along the $x$ direction.  Although the interferometer is nominally one dimensional along $x$, the Coriolis force causes transverse deflections so that at point C the upper arm $z$-position is $z_C= \tfrac{1}{2}\left|2\v{\Omega}\times\v{v}\right|T^2=\Omega_\text{or} (k_\text{eff}/m) T^2\sim 2~\text{cm}$, where once again $\Omega_\text{or}$ is the orbital angular velocity of the AGIS satellites.  For increased symmetry between the arms, the interferometer in Fig. \ref{Fig:FourPulseSequence} uses double-diffraction beamsplitters \cite{Lang} in which the upper and lower arms of the interferometer are given a positive and negative recoil kick, respectively.  In contrast to the typical single-diffraction beamsplitter where only one arm changes velocity, the double-diffraction beamsplitter results in symmetric Coriolis deflections which are needed to close the interferometer.

Multiple-pulse sequences such as the one in Fig. \ref{Fig:FourPulseSequence} also help reduce systematic errors in the instrument.  For example, by adjusting the time spacings between the pulses, the four-pulse sequence can be tuned to be insensitive to either uniform accelerations or gravity gradients and Coriolis effects \cite{BorisDiamonds, AGIS}.  This is useful for suppressing spurious noise from these sources.  The sequence shown in Fig. \ref{Fig:FourPulseSequence} uses a symmetric pulse spacing with pulses occurring at times $t_1=0$, $t_2=T$, $t_3=3T$, and $t_4=4T$.  As a result, this interferometer is insensitive to acceleration noise, but it remains sensitive to gravity gradients.

A five-pulse sequence ($\pi/2-\pi-\pi-\pi-\pi/2$) offers a number of advantages that make it a promising candidate for the AGIS-LEO mission.  Figure \ref{Fig:FivePulseSequence} shows an example of a symmetric five-pulse sequence with pulses at times $t_1=0$, $t_2=T$, $t_3=3T$, $t_4=5T$, and $t_5=6T$ and with variable pulse wavevectors $k_i$ along the $x$-direction assuming large momentum transfer (LMT) atom optics with magnitudes $k_1=k_\text{eff}$, $k_2=\tfrac{9}{4} k_\text{eff}$, $k_3=\tfrac{5}{2} k_\text{eff}$, $k_4=\tfrac{9}{4} k_\text{eff}$, and $k_5=2 k_\text{eff}$, where $\hbar k_\text{eff}=200\hbar k$ (see Sec. \ref{SubSec:AtomOptics}).  The three mirror pulses in this sequence result in a three-loop interferometer geometry along the $x$-direction, and these particular $t_i$ and $k_i$ values are chosen to simultaneously meet the constraints of interferometer closure in $x$ and $z$ and symmetry in time (see the detailed discussion in Sec. \ref{subsection:ErrorModel}).  For four- and five-pulse sequences with the same low-frequency detection limit, the $z$-deflection of the five-pulse interferometer is intrinsically smaller than that of the four-pulse interferometer since it goes through zero at the midpoint.  Smaller deflection is desirable since the atoms must stay within the beam waist of the interferometer lasers (see Sec. \ref{sec:AIBeams}).  In addition, the five-pulse sequence is insensitive to both acceleration noise and gravity gradient induced background noise which reduces the temperature requirements (kinematic constraints) of the atom source (see Sec. \ref{SubSec:ColdAtomSource}).

\subsection{The Gravitational Wave Phase Shift}

\label{Sec:gwphase}

The gravitational wave-induced phase difference between two five-pulse ($\pi/2-\pi-\pi-\pi-\pi/2$) atom interferometers separated by baseline distance $L$ along the direction of propagation of the laser light is
\be \Delta\phi_\text{GW}=8 k_\text{eff} h L\sin^4\!{\left(\omega T/2 \right)}\left(\frac{7+8 \cos{\omega T}}{2}\right)\sin{\theta_\text{GW}} \label{Eqn: GW phase}\ee
where $h$ is the gravitational wave strain for a gravitational wave of frequency $\omega$, $T$ is the interrogation time of the interferometer, and $k_\text{eff}$ is the effective wavevector of the AI beamsplitter.  Here $\theta_\text{GW}=\omega t$ is the phase of the gravitational wave at the time $t$ of the measurement.  This result follows from a fully relativistic phase shift calculation discussed in \cite{Dimopoulos:2006nk, GRAtom} applied to the five-pulse sequence described in Sec. \ref{Sec: AI}.

To maximize the strain sensitivity of the detector, the baseline $L$ should be made as large as possible.  This is a straightforward scaling to implement since only the laser light needs to travel the distance $L$ between the atom interferometers; each AI can remain relatively small, situated at the ends of the long baseline (see Fig. \ref{Fig:satelliteSchematic}).  From Eq. (\ref{Eqn: GW phase}), the interferometer is maximally sensitive to GW frequencies at and above the corner frequency $\omega_c=\tfrac{2}{T} \cos^{-1}{\!\sqrt{\tfrac{3}{8}}}\approx 2\pi\cdot 0.29/T$ at which point $\Delta\phi_\text{GW}=(125/16) k_\text{eff} h L\sin{\theta_\text{GW}}$.  For lower frequencies $\omega<\omega_c$, the $3~\text{dB}$ point occurs at $\omega_\text{3dB}=\tfrac{2}{T} \csc^{-1}{\![4/\sqrt{5}]}\approx 2\pi\cdot 0.19/T$ and sensitivity falls off as $\omega^{4}$.  At higher frequencies $\omega>\omega_c$, the envelope of the sensitivity curve is constant and the periodic anti-resonant frequencies that are present in (Eq. 1) can easily be avoided by varying $T$ \cite{AGIS}.

In order to observe a gravitational wave, the phase shift $\Delta\phi_\text{GW}$ must be sampled repeatedly as it oscillates in time due to the evolution of $\theta_\text{GW}$.  To avoid aliasing, the sampling rate $f_r$ must be at least twice the frequency of the gravitational wave.  The sampling rate can be increased (without decreasing $T$) by operating multiple concurrent interferometers using the same hardware \cite{AGIS}.  In these multiplexed interferometers, atom clouds with different velocities would be addressed via Doppler shifts of the laser light.

\subsection{Gravitational Wave Sensitivity}
\label{SubSec:GravitationalWaveSensitivity}

\begin{figure}
\includegraphics[width=7.0 in]{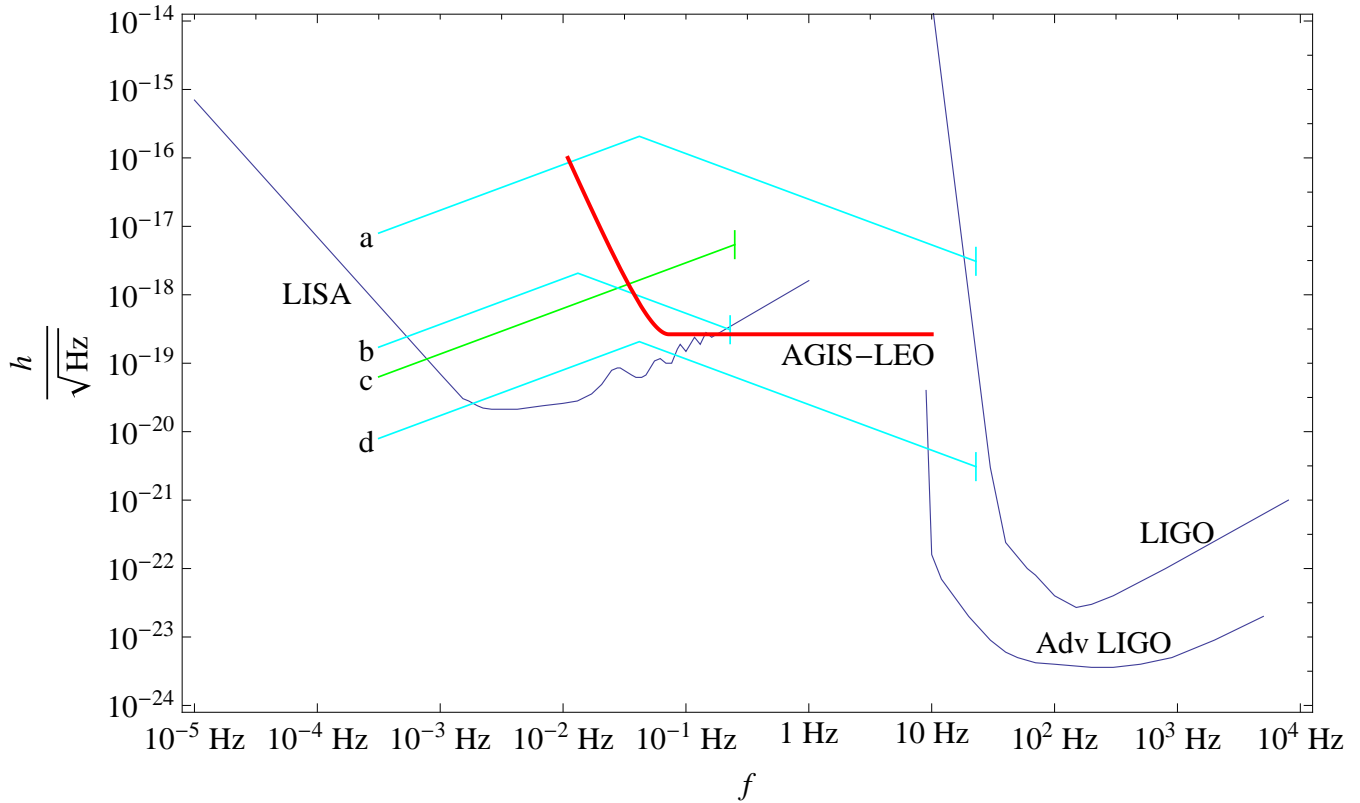}
\caption{ \label{Fig:sensitivityAGISLEO} The thick (red) curve shows the enveloped strain sensitivity versus gravitational wave frequency of a five-pulse interferometer sequence for a single arm of the proposed AGIS-LEO instrument.  Analogous plots for LIGO \cite{LIGOSensitivity}, Advanced LIGO \cite{Smith:2009cq}, and LISA \cite{Larson:1999we} are included for comparison. Curves a-d show gravitational wave source strengths after integrating over the lifetime of the source or one year, whichever is shorter: (a) represents inspirals of $10^3 M_{\odot}$, $1 M_{\odot}$ intermediate mass black hole binaries at 10 kpc, (b) inspirals of $10^5 M_{\odot}$, $1 M_{\odot}$ massive black hole binaries at 10 Mpc, (c) white dwarf binaries at 10 kpc, and (d) inspirals of $10^3 M_{\odot}$, $1 M_{\odot}$ intermediate mass black hole binaries at 10 Mpc.}
\end{figure}

The phase sensitivity of an atom interferometer is limited by quantum projection noise $\delta\phi= 1/\text{SNR}=1/\sqrt{N_a}$, where $N_a$ is the number of detected atoms per second.  For shot-noise limited atom detection, the resulting strain sensitivity to gravitational waves becomes $h=\tfrac{\delta\phi}{8 k_\text{eff} L}\csc^4\!{\left(\omega T/2 \right)}\left(\tfrac{2}{7+8 \cos{\omega T}}\right)$.  Figure \ref{Fig:sensitivityAGISLEO} shows the AGIS-LEO mission strain sensitivity (enveloped) for a single pair of satellites using the five-pulse interferometer sequence from Fig. \ref{Fig:FivePulseSequence} with $\hbar k_\text{eff}=200\hbar k$ beamsplitters, an $L=30~\text{km}$ baseline, a $T=4~\text{s}$ interrogation time, and a phase sensitivity of $\delta\phi=10^{-4}~\text{rad}/\sqrt{\text{Hz}}$.  Although the sensitivity spectrum contains periodic nulls corresponding to frequencies at which $\Delta\phi_\text{GW}=0$, in practice the entire strain sensitivity envelope shown in Fig. \ref{Fig:sensitivityAGISLEO} can be accessed by scanning $T$ over a small range on repeated interferometer sequences.

\begin{figure}
\includegraphics[width=7.0 in]{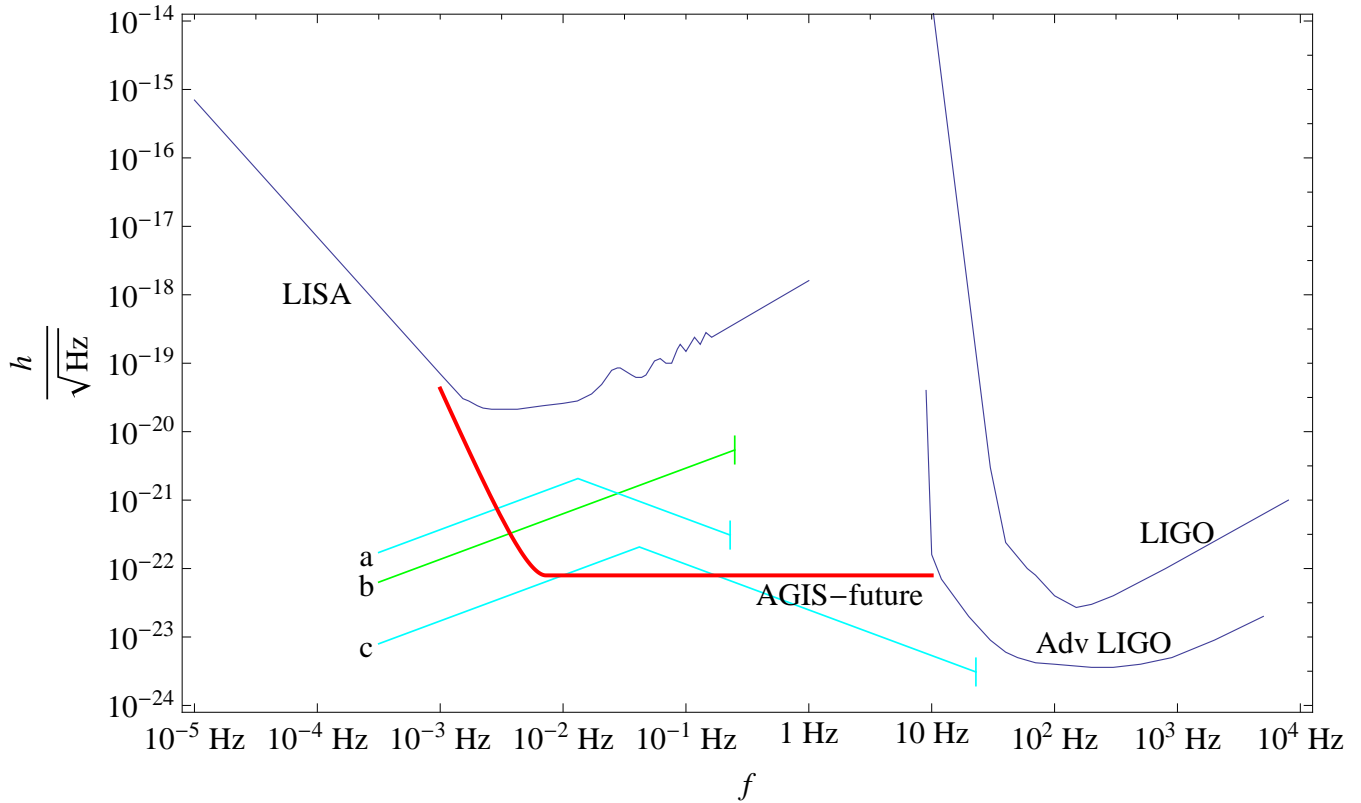}
\caption{ \label{Fig:sensitivity2} The thick (red) curve shows the enveloped strain sensitivity versus gravitational wave frequency of a five-pulse interferometer sequence for a single arm of a sensor that could be realized in the future. Analogous plots for LIGO \cite{LIGOSensitivity}, Advanced LIGO \cite{Smith:2009cq}, and LISA\cite{Larson:1999we} are included for comparison. Curves a-c show gravitational wave source strengths after integrating over the lifetime of the source or one year, whichever is shorter: (a) represents inspirals of $10^5 M_{\odot}$, $1 M_{\odot}$ massive black hole binaries at 10 Gpc, (b) white dwarf binaries at 10 Mpc, and (c) inspirals of $10^3 M_{\odot}$, $1 M_{\odot}$ intermediate mass black hole binaries at 10 Gpc. It should be noted that the stochastic background of white dwarf binaries will likely overwhelm all but the strongest of these sources at frequencies less than $\sim 0.5$ Hz (see Fig. \ref{Fig:stochastic}).  Note also that this sensitivity curve assumes atom shot noise limited detection and does not include limits that could be imposed by systematic effects present in the future apparatus.}
\end{figure}

Many of the parameters that set the instrument sensitivity can eventually be improved as AI technology continues to evolve.  Figure \ref{Fig:sensitivity2} shows the gravitational wave sensitivity that would be possible with a more advanced version of the AGIS instrument.  This future sensitivity curve assumes the same five-pulse interferometer sequence and $\hbar k_\text{eff}=200\hbar k$ atom optics, but with an $L=10^4~\text{km}$ baseline, a $T=40~\text{s}$ interrogation time, and an improved phase sensitivity of $\delta\phi=10^{-5} \; \text{rad}  / \sqrt{\text{Hz}}$.  We note that this sensitivity curve does not include limits that might be imposed by the systematic effects relevant to such a future apparatus, since its purpose is simply to illustrate the potential for atomic gravitational wave detectors to improve as technology advances.  The large baseline of this future instrument would require a solar orbit and is not addressed here.

\subsection{Satellite Configuration}
\label{Sec:SatelliteConfiguration}
\subsubsection{Two-Satellite Configurations}
\label{Sec:Two-Sat}

\begin{figure}
\begin{center}
\includegraphics[width=5.0 in]{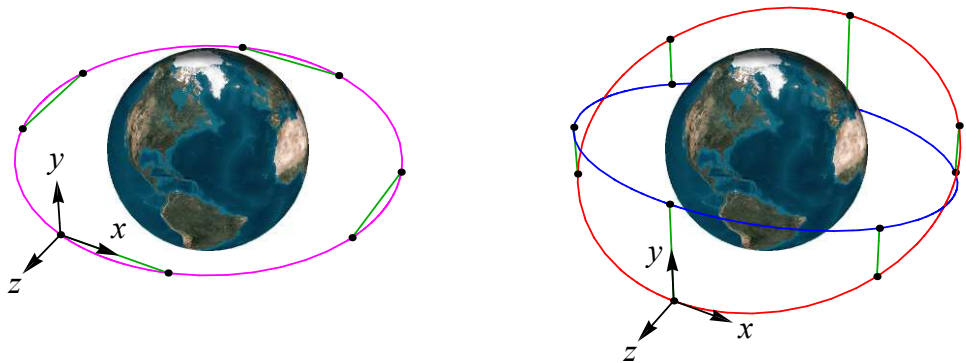}
\caption{ \label{Fig:greatCircles} Conceptual orbit diagrams for two possible AGIS-LEO two-satellite configurations, not to scale.  In each figure, the locations of the satellites (black dots) are shown at quarter-intervals (left) or sixth-intervals (right) of the orbital period, connected by a green line corresponding to the interferometer baseline axis of the gravitational wave detector at that instant.  \emph{Left}: A leader-follower (LF) configuration in which both satellites follow circular orbits with the same altitude and angle of inclination, but with different angles of true anomaly (i.e. phases).  In this formation, the baseline's length does not change, but it rotates constantly as the satellites orbit the Earth.  Unless otherwise specified, the leader-follower orbit is assumed in this paper.  \emph{Right}: An inclined-great-circles (IGC) configuration in which both satellites are in circular orbits of the same radius, but with different angles of inclination.  If the true anomaly difference between the satellites were zero, then the baseline axis would not rotate, but it would flip twice per period and it's length would oscillate.  Practically, however, it is necessary to add some anomaly to ensure that the satellites do not collide.  In this case, one satellite appears to move in an ellipse centered on the other.  The eccentricity of this ellipse is determined by the ratio between the anomaly and the inclination, allowing continuous tuning from a configuration that favors nearly constant orientations at the cost of length variations to a configuration with invariant lengths and constant rotation rates.  Note the difference in definitions of the coordinate axes in the satellite frames between the LF and IGC configurations.  In the LF configuration (as in Fig \ref{Fig:satelliteSchematic}), the $x$-direction is fixed to the rotating baseline and the $y$-direction is normal to the satellites' mutual orbital plane.  In the IGC configuration, the baseline lies on the $y$-axis and the $x$-axis is chosen to be tangential to the satellite's orbital curve.}
\end{center}
\end{figure}

As shown in the preceding sections, gravitational waves can be detected with a pair of atom interferometers manipulated by two satellites separated by a baseline of length $L$.  Ideally, the satellites would be inertial and the baseline length would be constant.  Instead, many practical issues must be addressed for the non-inertial low Earth orbits we discuss here.  The simplest two-satellite configuration is the circular leader-follower (LF) configuration (see Fig. \ref{Fig:greatCircles}), which yields baselines of constant length, but also with constant orbital rotation rate, $\Omega_\text{or}$.  While this rotation bias yields off-axis Coriolis deflections (as discussed in Sec. \ref{Sec: AI} and Figs. \ref{Fig:FourPulseSequence} and \ref{Fig:FivePulseSequence}), the LF configuration provides a simple starting point for our analysis.  Thus, an equatorial, circular, two-satellite, LF orbit will be assumed throughout this paper unless stated otherwise.

However, it should be noted that other orbital configurations exist for two satellites that have some advantages over the simple LF configuration.  For instance, the inclined-great-circles (IGC) configuration shown on the right in Fig. \ref{Fig:greatCircles} reduces the impact of rotation bias at the cost of a variable baseline length.  While any orbit pair at equal altitude requires the satellites to circle each other once per period, this configuration can concentrate most of this rotation at the orbit crossing points.  During the rest of the orbit, the baseline deviates little from one of two antiparallel orientations.  For instance, at an orbital altitude of $1000~\text{km}$ (105 minute period), an IGC orbit with a maximum separation of $30~\text{km}$ and a minimum separation of $100~\text{m}$ exhibits sub-degree angular deviations during two 45-minute windows centered at the points of maximum separation.

\begin{figure}
\begin{center}
\includegraphics[width=5.0 in]{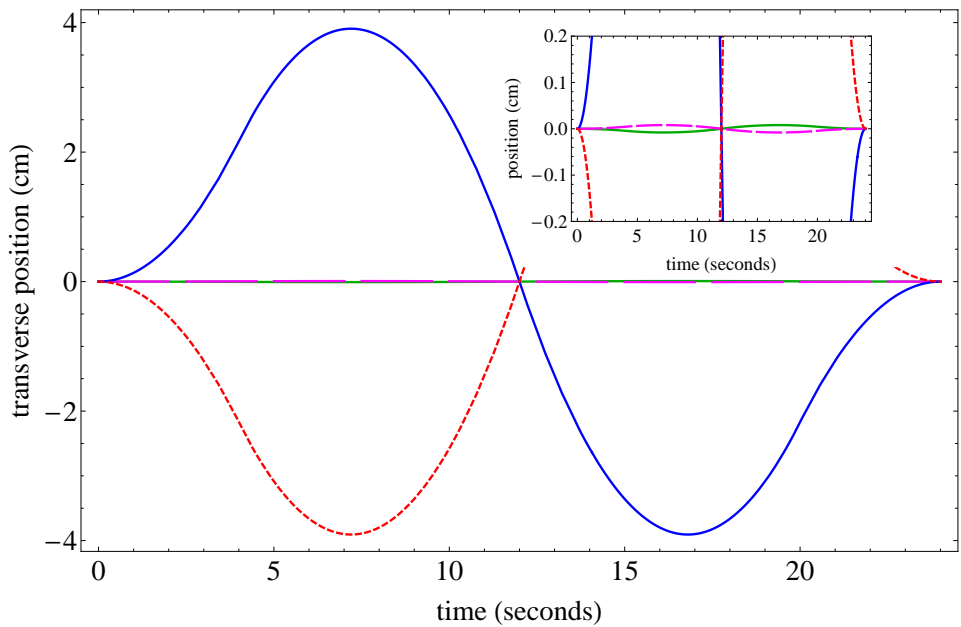}
\caption{ \label{Fig:TransverseMotionComparsion} Coriolis deflection from the interferometer axis for inclined-great-circles and leader-follower orbits. Leader-follower deflection is shown in solid (blue) and dotted (red) lines with a maximum transverse separation of $8~\text{cm}$ between the two arms of the interferometer, while the inclined-great-circles deflection is shown in solid (green) and dashed (purple) lines with a maximum separation of $160~\mu \text{m}$. The inset shows a close-up view in order to detail the inclined-great-circles deflection.}
\end{center}
\end{figure}

The potential benefits of the IGC configuration are substantial.  Most important, the small angular deviations decrease rotation bias and Coriolis deflections (see Fig. \ref{Fig:TransverseMotionComparsion}), relaxing many of the constraints based on $\Omega_\text{or}$ in the LF configuration.  Further improvements are then possible with star-tracker-based satellite pointing stabilization and active servoing of the beam delivery optics to compensate for residual rotation \cite{varenna}.  Finally, if the baseline axis points toward the Sun, this configuration makes a sunshield more practical, since it would only have to protect the interferometer region along its transverse extent rather than its full length.  (As will be discussed in Sec. \ref{sec:orbAlt}, the Earth's shadow could also be used for sun shielding, yielding about 35 minutes of uptime per period at $1000~\text{km}$ of altitude.  This corresponds well with the 45-minute window of small angular deviations discussed previously.)

The variable baseline length in an IGC configuration implies that the atoms from the two satellites have a non-zero relative velocity along the interferometer axis.  This requires that the lasers be phase-chirped to account for the Doppler shift caused by this changing relative velocity over the duration of the interferometer.  Additionally, the atoms released from the two satellites will have different respective Doppler shifts.  Thus, two different sets of frequencies will be needed to address a pair of interferometers in a differential configuration (with one interferometer released from each satellite).  The two frequency sets can be realized by electro-optically generating multiple sidebands from a common source laser.

The relative velocity between the two interferometers is advantageous for Doppler-based multiplexing of multiple interferometers during the same time window. The relative velocity between the atoms and the satellite depends on $\Delta T$, the time since the start of each interferometer.  Thus, each cloud is naturally tagged with a different frequency which allows for easy individual addressability.

\subsubsection{Three-Satellite Configurations}
\label{Sec:Three-Sat}

\begin{figure}
\begin{center}
\includegraphics[width=5.0 in]{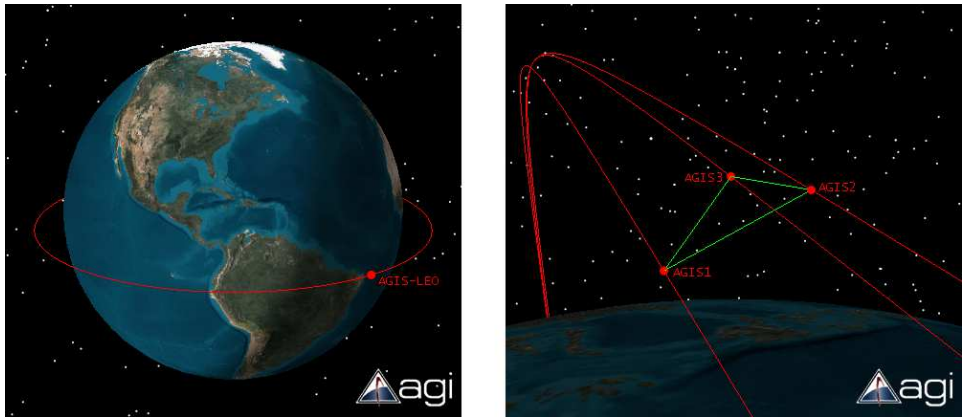}
\caption{ \label{Fig:AGIS-SatelliteOrbits} AGIS-LEO three-satellite orbit simulations.  \emph{Left}: Equatorial orbit of the entire AGIS-LEO constellation at an altitude of $1000~\text{km}$.  \emph{Right}: Closer view of three AGIS satellites in an equilateral triangular formation.  The simulations were done using the STK software package from AGI.}
\end{center}
\end{figure}

While a single pair of atom interferometers along a common baseline is sufficient for detecting gravitational waves, a multiple-arm configuration with several AI pairs offers many potential advantages, which we will discuss shortly.  Our three-arm proposal consists of three satellites maintaining constant separations of $\sim 30~\text{km}$ in formation flight (Fig. \ref{Fig:AGIS-SatelliteOrbits}).  Such a formation can be pictured as a chief point on a circular Earth-orbit, with deputy satellites in circular orbits in a Hill frame around the chief point.  It has been shown that such orbits exist with low station-keeping costs even when one accounts for the $J_2$ oblateness of the Earth \cite{ref:sneeuw}.  Lasers located inside each satellite send light to both of the other satellites, forming a set of three arms consisting of counter-propagating light beams.  Each arm has an atom interferometer at both ends - near, but outside of, the satellites (Fig. \ref{Fig:satelliteSchematic}).  The pair of atom interferometers along a given arm is manipulated by the counter-propagating light beams of that arm, and the phase difference between these AIs due to a gravitational wave is given by Eq. (\ref{Eqn: GW phase}).  As a result, each arm of the instrument acts as an independent gravitational wave detector.

\begin{figure}
\includegraphics[width=7.0 in]{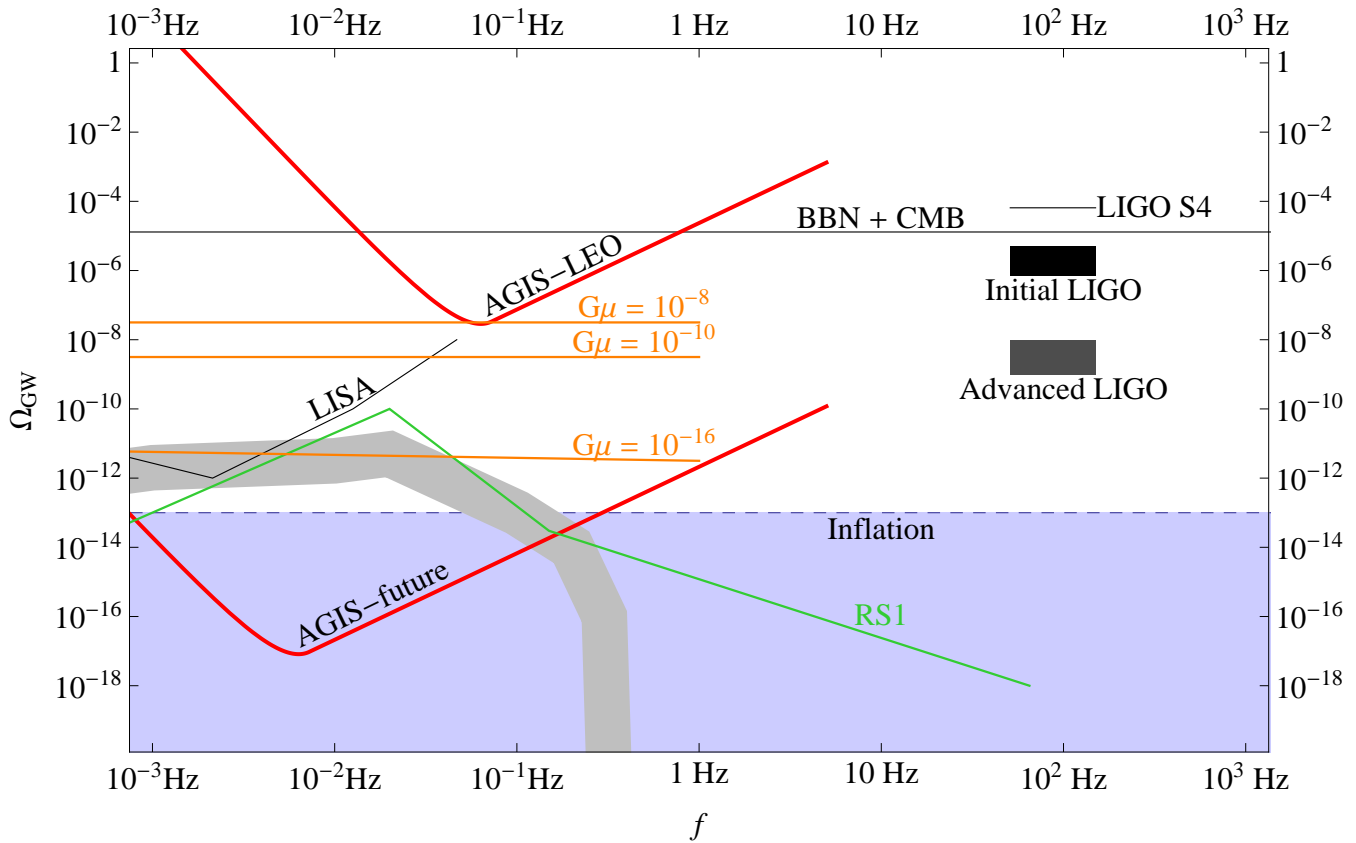}
\caption{ \label{Fig:stochastic} The 95\% confidence limits on the sensitivity of the instrument in terms of gravity wave energy density, $\Omega_{GW}$, for present and future parameters are plotted as thick (red) curves. These sensitivities are for a five-pulse interferometer sequence. The limit from the LIGO Science Run 4 and the projected limits from initial and advanced LIGO are shown \cite{Abbott:2006zx}. The limits from models of big-bang nucleosynthesis (BBN) \cite{Allen:1996vm} and the cosmic microwave background (CMB) \cite{Smith:2006nka} apply to the integral of the stochastic gravitational wave background over frequency. The possible region of gravitational waves produced by a period of inflation (not including reheating) is shown. The upper limit on this region is set by the COBE bound \cite{Allen:1994xz}. The gray band shows a prediction for the stochastic gravitational wave background from extragalactic white dwarf binaries; its width shows an expected error \cite{Farmer:2003pa}. The (green) curve labeled RS1 corresponds to an example spectrum of gravity waves from a TeV scale phase transition, in this case from the RS1 spacetime geometry \cite{Randall:2006py}.  The (orange) lines labeled with $G \mu$ correspond to one prediction for a network of cosmic strings with tensions $G \mu = 10^{-8}$, $G \mu = 10^{-10}$ and $10^{-16}$ (with $\alpha = 0.1$ and $\gamma = 50$) \cite{DePies:2007bm}.  Note that these cosmic string estimates have large uncertainties and may be optimistic assumptions.}
\end{figure}

The desire for multiple, independent GW detectors in the AGIS instrument is motivated by one of our primary science goals: the detection of stochastic GWs.  While a coherent GW signal, such as one from a binary merger, can be detected by a single arm of AGIS, stochastic GWs can be detected only by studying correlations in the signals from at least two independent GW detectors.  In Fig. \ref{Fig:stochastic}, the sensitivities of the AGIS-LEO and future instruments are plotted against expected stochastic backgrounds. Following the analysis in \cite{AGIS}, we display the sensitivities as 95\% confidence limits on the gravity wave energy density, $\Omega_{GW}$. To account for the effect of the satellite configuration on the instrument's sensitivity to stochastic sources, the calculation includes the appropriate overlap function value $\gamma = \pi$ for an equilateral triangle geometry \cite{Christensen:1992wi}.  As indicated by the sensitivity of the future-technology curve, sensors using AIs have the potential to detect stochastic GWs from cosmic strings and, depending on the exact frequency cutoff of the white dwarf background, possibly even inflation and TeV scale phase transitions.

There are also technical advantages to the AGIS-LEO three satellite configuration over a single arm design.  Additional suppression of laser phase noise and vibration noise is possible by combining the signals from multiple arms.  Noise that originates from a particular satellite can have a significant common component for the two arms serviced by that satellite.  Comparison of the signals from multiple arms can be used to subtract this common noise, allowing for additional noise cancellation in the event that the single arm differential suppression is not sufficient.  Finally, the use of three independent arms provides redundancy and improves uptime.  A final determination of the AGIS-LEO orbital configuration will require a more thorough analysis of the costs and benefits outlined here.

\section{Instrument Overview}

A single phase measurement in an atom interferometer consists of three steps: atom cloud preparation, interferometer pulse sequence, and detection. In order to attain the $10^{-4} \; \text{rad}  / \sqrt{\text{Hz}}$ phase sensitivity, the instrument must be designed to provide, maintain, and measure a cold atom flux of $10^8$ atoms/s. Instrument parameters resulting from the discussion below are summarized in Table \ref{tab:AtomSource}.

\begin{table}
\setlength{\tabcolsep}{5pt}
\begin{tabular}{lll}
Parameter & Specification & Location in Text\\
\hline
\hline
Baseline & $30$ km & Gravitational Wave Sensitivity (Sec. \ref{SubSec:GravitationalWaveSensitivity})\\
Atom Flux & $ 10^8 \; \frac{\text{atoms}}{\text{s}}$ & Cold Atom Source (Sec. \ref{SubSec:ColdAtomSource})\\
Cloud Temperature & $\sim 100 \; \text{pK}$ & Cold Atom Source and Wavefront (Sec. \ref{SubSec:ColdAtomSource} and \ref{SubSubSec:ShotToShot})\\
Cooling Laser Powers & $\sim 100 \; \text{mW}$ & Cold Atom Source (Sec. \ref{SubSec:ColdAtomSource})\\
Interferometer Sequence & Multiple loop & Atom Interferometer Pulse Sequence (Sec. \ref{Sec: AI})\\
Momentum Separation of Interferometer Arms & $\sim 200 \hbar k$ & Atom Optics (Sec. \ref{SubSec:AtomOptics})\\
\end{tabular}
\caption{Instrument parameters summary. For interferometer laser requirements, see Table \ref{tab:InterferometerBeams}.}
\label{tab:AtomSource}
\end{table}

\subsection{Cold Atom Source}
\label{SubSec:ColdAtomSource}

The suppression of velocity-dependent backgrounds requires the atoms to be cold, with RMS velocity widths as small as $\sim 100 \; \mu \text{m/s}$, corresponding to cloud temperatures of $\sim 100 \; \text{pK}$.  Cold atom clouds with $10^8$ to $10^{10}$ atoms are readily produced using modern laser cooling techniques \cite{ref:Metcalf}.  Furthermore, it has been demonstrated that the hardware necessary for these standard techniques is simple and can be made robust and compact - enough to perform well in drop towers and on airplanes \cite{vanZoest,philippePlane,varenna}.  It is possible to make cold atom sources with fluxes of $10^{10} \text{ atoms/s}$ that fit in a volume of less than $20 \; \text{cm}^3$ and require $\sim 100 \; \text{mW}$ of laser power \cite{conroy}.

The required $\sim 100 \; \mu \text{m/s}$ wide cloud can be extracted from a large ($\gtrsim 10^{10}$ atom) $\mu$K-temperature thermal cloud by applying a velocity-selective cut using Doppler sensitive two-photon transitions \cite{ref:raman}.  Alternatively, the necessary velocity spread can be achieved by adiabatically expanding a cold atom ensemble in a harmonic trap or by implementing delta kick cooling \cite{ref:deltakick}, both of which are aided by being in microgravity, or by cooling the atoms in an optical lattice \cite{bernier}.  Adiabatic expansion of a cold atom ensemble in an all-optical trap, which requires only simple cavities and low power lasers, can be realized with a setup similar to that described in \cite{ref:philippeoptical}.  In any of these cases, low densities are desirable to mitigate possible systematic noise sources associated with cold collisions.  If necessary, an array of multiple atom sources can be placed in each satellite in order to meet the atom flux requirement.

After preparation, the cold atom ensemble must be positioned outside the satellite in the interferometer region (see Figs. \ref{Fig:satelliteSchematic} and \ref{Fig:AGIS-Telescope}).  This can be accomplished using laser light along the primary interferometer beam axis.  Counter-propagating light from two satellites is used to form an optical standing wave, trapping the cold atoms.  Such an optical lattice trap can then shuttle the atom cloud out to the appropriate initial position \cite{ref:denschlag}.

In leader-follower low Earth orbits, positioning the atoms outside the satellites must account for significant Coriolis forces.  For example, moving an atom cloud $10~\text{m}$ from the satellite in a time of $1~\text{s}$ can lead to transverse accelerations of $\left|\v{v}\times\v{\Omega}\right|\sim 10^{-3}g$.  Using a red--detuned optical lattice can provide transverse confinement during cloud positioning, but this may require small lattice beam diameters, potentially limiting the baseline due to laser light diffraction effects.  To avoid this baseline constraint, a more tightly focused beam may be sent out from each satellite in addition to the larger diameter optical lattice, acting as a source of transverse confinement for the atoms near the satellite from which the small beam originates.

\subsection{Atom Optics}
\label{SubSec:AtomOptics}

Conventional Raman atom optics transfer momentum $\hbar k_\text{eff}=2\hbar k$ to the atom at each beamsplitter \cite{kasevichChuAI}.  To achieve the desired sensitivity, it is necessary to increase $k_{\text{eff}}$ using large momentum transfer (LMT) beamsplitters.  An $N$th order LMT beamsplitter (corresponding to an $\hbar k_\text{eff}=2N\hbar k$ momentum separation between the arms of an interferometer) yields an $N$-fold enhancement of the phase difference in Eq. (\ref{Eqn: GW phase}).  LMT beamsplitters can be achieved by adiabatically accelerating the atoms with an optical lattice \cite{ref:kovachy, ref:denschlag, ref:clade, ref:muellerlattice} or by applying sequential two-photon Bragg pulses.  In addition to the desired stimulated transitions, the atoms can undergo unwanted spontaneous two-photon transitions when in the presence of the laser field, consisting of the stimulated absorption of a photon from the laser field followed by the spontaneous emission of a photon in a random direction.  Atoms that undergo spontaneous emission are lost and do not contribute to the interferometer signal, so this effect must be minimized.  The spontaneous emission rate can be written as $R_{\text{sc}} = \frac{2 \Omega_{\text{st}}^2}{\Gamma (I/I_{\text{sat}})}$, where $\Omega_{\text{st}} = \frac{\Gamma^2 (I/I_{\text{sat}})}{4 \Delta}$ is the Rabi frequency of the desired stimulated two-photon transitions, $I$ is the laser intensity, $\Gamma$ is the spontaneous decay rate of the excited state, $\Delta$ is the laser detuning, and $I_{\text{sat}}=\frac{\hbar \omega\Gamma}{2\sigma_0}$ is the saturation intensity of the atomic transition with resonant cross section $\sigma_0\sim \lambda^2$ and frequency $\omega=2\pi c/\lambda$.  The time $t_N$ needed for an $N$th order LMT beamsplitter is given by $t_N \sim N \frac{\pi}{\Omega_{\text{st}}}$.  In order for spontaneous emission not to degrade the signal significantly for $q$ concurrent interferometers, we therefore require that $q N \frac{\pi} {\Omega_{\text{st}}} \lesssim \frac{1}{R_{\text{sc}}}$.  For example, for rubidium atoms ($I_{\text{sat}} \approx 2.5 \; \frac{\text{mW}}{\text{cm}^2}$ and $\Gamma \approx 2\pi\cdot 6.066~\text{MHz}$) we can meet this requirement for $N \sim 100$ LMT beam splitters and $q \sim 10$ concurrent interferometers by choosing $\Omega_{\text{st}} \sim 2\pi \times 0.4 \; \text{kHz}$ and laser powers of $\sim 1 \; \text{W}$ with a beam waist of $10 \; \text{cm}$.

In addition to minimizing spontaneous emission, we must ensure that noise in the differential phase between interferometers due to AC Stark shifts \cite{ref:Metcalf} (including those due to off-resonant two-photon transitions) is below $10^{-4} \; \text{rad}  / \sqrt{\text{Hz}}$.  This can be achieved with the parameters given above by stabilizing fluctuations in the laser intensity at the $\sim 0.01 \%$ level.  Off-resonant shifts are discussed in detail in \cite{ref:kovachy}.

When multiple concurrent Doppler-multiplexed interferometers are used, for certain parameters it may be optimal to choose the launch momentum difference between successive interferometers to be smaller than the momentum splitting between the arms of a single interferometer.  If this is the case, then there will be instances when two arms of two different interferometers will cross each other in momentum space.  The existence of such crossings will require only minor alterations to the pulse sequence to avoid interactions with the undesired arm.

\subsection{Detection}
\label{SubSec:Detection}

A normalized phase measurement at the end of a pulse sequence consists of counting the number of atoms at each output port of the interferometer.  Since the interferometry takes place outside the satellite, this counting must be done remotely.  Fluorescence imaging, commonly used in laser cooling and trapping experiments, is an inherently remote state-sensitive atom-counting technique.  A fluorescence-based detection scheme can be implemented with CCD cameras on the satellites and the same telescopes used to enlarge the primary atom interferometry beams (see Sec. \ref{subsec:TelescopeDesign}).  For instance, a detection beam tuned to resonance with an optically allowed atomic transition can be generated and enlarged to a $10 \; \text{cm}$ waist with the telescope on one satellite.  Backscattered fluorescent light can be collected using the same telescope.

The details of such a scheme are as follows.  Prior to detection, atoms exiting a port of the atom interferometer can be transferred from the interferometer atomic state (one of the ground state hyperfine levels) to the imaging state (the other ground state hyperfine level) using a Doppler sensitive two-photon Raman transition.  Here the Doppler sensitivity of the transition is used to distinguish the output ports of the interferometer, since the atoms in the different output ports have different longitudinal velocities (see for instance, Fig. \ref{Fig:FivePulseSequence}).  A subsequent short detection pulse ($\sim 1$ ms) (tuned to be resonant with the transition between the imaging state and an optical excited state) scatters fluorescent photons.  A fraction of these photons is collected by the imaging optics and focused onto a CCD detector.  With $N_a \sim 10^8$ atoms in the cloud, a phase sensitivity of $\sim 10^{-4}~\text{rad}$ requires the ability to measure changes in atom number at the  $\sqrt{N_a} \sim 10^4$ level (the atom shot-noise limit).  This limit can be achieved by scattering enough photons from the atomic ensemble so that at least 1 photon per atom is detected.  The required number of scattered photons is determined by the solid angle subtended by the light collecting optics, which for a $30~\text{cm}$ primary telescope mirror at a distance of $10~\text{m}$ from the atom cloud is $\Delta\Omega\sim 6 \times 10^{-5}~\text{sr}$.  For the above parameters, the detection laser must have $I/I_\text{sat} \sim 10$ to achieve 1 photon per atom in $1~\text{ms}$. The outputs from multiple Doppler-multiplexed interferometers run concurrently can be detected with a sequence of appropriately tuned velocity-selective Raman pulses.  Finally, the fact that the detection is spatially resolved allows for characterization of possible optical wavefront curvature errors in the sensor (see Sec. \ref{SubSubSec:WavefrontMitigation}).

Having shown that it is possible to produce, maintain and detect the necessary cold atom flux, we now consider the noise sources that may compete with gravitational waves at this level of phase sensitivity. These sources fall into two general categories: noise of the laser beam wavefronts and noise from the environment. The former affects the atoms' reference frame while the latter affects the atoms' motion relative to that reference frame.

\section{Atom Interferometer Laser Beam Considerations}
\label{sec:AIBeams}

The laser wavefronts form the reference frame against which the motion of the atoms is compared. Thus, it is important that any non-common noise in the wavefronts that is imprinted on the interferometers falls below the phase sensitivity limit in the frequency band of interest. In this section, we discuss telescope design, wavefront aberrations, and laser phase noise as they relate to this requirement. For a summary of the resulting interferometer laser pulse parameters in particular, see Table \ref{tab:InterferometerBeams}.

\begin{table}
\setlength{\tabcolsep}{5pt}
\begin{tabular}{lll}
Parameter & Specification & Location in Text\\
\hline
\hline
Power & $\sim 1$ W & Atom Optics (Sec. \ref{SubSec:AtomOptics})\\
Waist & $\sim 10$ cm & Atom Optics and Laser Pointing Angle Jitter (Secs. \ref{SubSec:AtomOptics} and \ref{subsec:PointingJitter})\\
Intensity Stability & $< 0.01 \%$ & Atom Optics (Sec. \ref{SubSec:AtomOptics})\\
Wavefront & $< \frac{\lambda}{1000}$ & Wavefront Aberrations (Sec. \ref{SubSec:WavefrontAberrations})\\
Phase Noise & $< -100 \;\frac{\text{dBc}}{\text{Hz}}$ & Laser Phase Noise (Sec. \ref{SubSec:LaserPhaseNoise})\\
Fractional Frequency Stability & $ < 10^{-11}$ & Laser Phase Noise (Sec. \ref{SubSec:LaserPhaseNoise})\\
\end{tabular}
\caption{Interferometer laser pulse requirements.}
\label{tab:InterferometerBeams}
\end{table}

\subsection{Telescope Design}
\label{subsec:TelescopeDesign}

\begin{figure}
\begin{center}
\includegraphics[width=5.0 in]{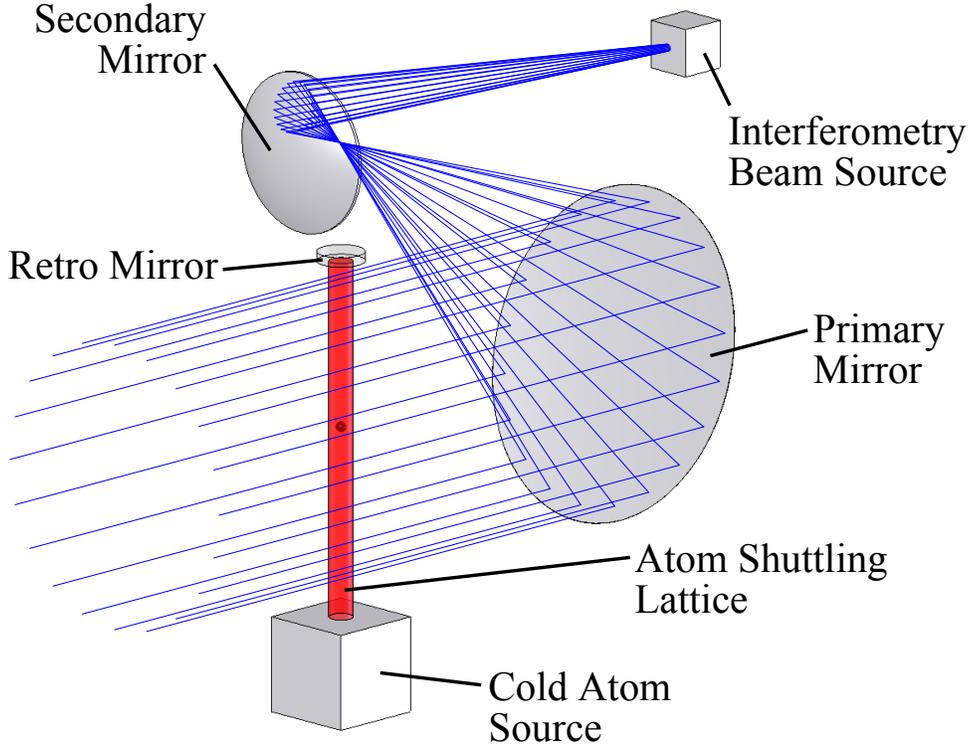}
\caption{ \label{Fig:AGIS-Telescope} Conceptual diagram of a single AGIS telescope as a $30 \; \text{cm}$ off-axis Gregorian system.  Note the intermediate focus of the Gregorian telescope, which allows for the use of pinhole spatial filtering to remove wavefront aberrations from upstream optics and lasers.  Also shown approximately to scale is the cold atom source and a lattice beam that could be used to shuttle the atoms into the middle of the primary interferometry beam.  The primary beam line can in turn be used to shuttle the atoms away from the satellite prior to an interferometer sequence.}
\end{center}
\end{figure}

To generate interferometry beams with Rayleigh ranges equal to half our satellite separations (see Sect. \ref{subsec:PointingJitter}), we will need telescopes to magnify the beam waists to $5-10~\text{cm}$.  These beam waists should also be large compared to the typical sizes of the atom clouds, $\sim 1~\text{cm}$ FWHM.  The telescopes serve multiple purposes in the instrument: their beam line can be used to shuttle atoms into the interferometry region before a measurement (Fig. \ref{Fig:satelliteSchematic}), they magnify the primary atom optics beams (Sec. \ref{SubSec:AtomOptics}), and they form a part of the detection imaging system (Sec. \ref{SubSec:Detection}).  As discussed in the remainder of this section, the atom optics beams place the most stringent requirements on the telescope design.  Each beam is used in two atom interferometers, one a few meters outside the laser's originating satellite, and the other $\sim 30~\text{km}$ away.  Thus, the telescope's design must account for both small-scale imperfections and large-scale aberrations. For instance, the secondary mirror must not obstruct the path of the beam and the primary mirror must be large enough to avoid aperture diffraction effects.  A $\sim 30~\text{cm}$ diameter off-axis Schiefspiegler or off-axis Gregorian telescope could be capable of meeting these requirements \cite{LISA_MissionConcept,COBE_DIRBE}.  A Gregorian telescope has the additional advantage that its real intermediate focus could be used for pinhole spatial filtering, which would eliminate wavefront errors from all optics and lasers before the primary mirror (see Fig. \ref{Fig:AGIS-Telescope}).

\subsection{Wavefront Aberrations}
\label{SubSec:WavefrontAberrations}

Wavefront aberration of the interferometer laser beams is a potential source of noise.  Since the phase of the laser is imprinted on the atom during each light pulse, any transverse phase variation results in a phase shift that depends on the transverse position of the atom with respect to the laser beam spatial profile.  This effect can result in phase noise (i) due to satellite transverse position noise during the interferometer pulse sequence, (ii) due to fluctuations in the mode itself induced by temporal variations in the optics path or (iii) due to jitter in initial atom cloud position or width from shot to shot.  The effect is in general not common-mode suppressed since wavefront perturbations with high transverse spatial frequency diffract out of the beam over distances smaller than the satellite separation.  The perturbations become common-mode suppressed at transverse wavelengths $\lambda_t$ longer than the corner wavelength $\lambda_c=\sqrt{L \lambda}\sim 17~\text{cm}\left(\frac{L}{30~\text{km}}\right)^{1/2}$, where $\lambda_c$ is defined as the transverse spatial length scale which has a Rayleigh range of $L$. In addition, since the in the simplest case the interferometer phase measurement is determined by the total number of atoms in a particular state, aberrations with transverse wavelengths smaller than the size of the cloud can be suppressed by spatially averaging over the atom cloud.  However, by performing this average, the phase measurement becomes sensitive to changes in the atom cloud size from shot to shot, a noise source which we also discuss below.

\subsubsection{Wavefront Aberration Phase Shift Calculation}
\label{SubSubSec:WavefrontCalculation}

The phase shift imparted to the atoms as a result of non-uniform transverse spatial phase of the laser wavefronts can be calculated by summing the local phase $\phi(x, z, t)$ imparted to the atom at each of the light pulses:
\be \Phi_t=\sum_{j=1}^{n} \phi(x_j, z_j, t_j) = \sum_{j=1}^{n}\theta(x_j, k_t, t_j) \sin\!{[k_t(z_j-\delta z(t_j))+\Theta_t]}\label{Eqn: Transverse Phase Calc}\ee
Here the position of the atom is referenced to the coordinate system shown in Fig. \ref{Fig:satelliteSchematic}, where $x_j$ is the longitudinal position of the atom along the direction of $k_\text{eff}$ at the time $t_j$ of the $j$th pulse and $z_j$ is the transverse position at time $t_j$ (for simplicity we suppress the transverse coordinate $y_j$).  The sum is taken over the $n$ light pulses of the interferometer sequence; in the cases considered here $n=5$ and the coordinates $(x_j, z_j, t_j)$ are taken to be those shown in Fig. \ref{Fig:FivePulseSequence}.  For simplicity, we consider the response to a single Fourier component of amplitude $\theta(x, k_t, t)$ and transverse spatial wavevector $k_t=2\pi/\lambda_t$.  The general phase $\Theta_t$ allows modeling of both quadratures.  The time-dependent reference position $\delta z(t)$ accounts for vibrational noise that affects the relative transverse position of the atom and the satellite.  In the paraxial limit, the amplitude of each Fourier component evolves as the beam propagates along the $x$ direction according to the paraxial propagator given by $\theta(x, k_t, t)=\theta(k_t, t)\exp{\left[-\imagI\tfrac{k_t^2}{2k}(x-x_0)\right]}$ where $x_0$ is the reference plane at which $\theta(k_t, t)$ is known.  Finally, in order to model time-dependent wavefront aberrations, we decompose each Fourier coefficient into a static and a dynamic component: $\theta(k_t, t)\equiv \theta_0(k_t)+ \delta\theta(k_t, t)$.  Here the term $\delta\theta(k_t, t)$ represents an explicitly time-dependent variation of the amplitude of the aberration with transverse wavevector $k_t$.

The phase shift $\Phi_t$ given by Eq. (\ref{Eqn: Transverse Phase Calc}) depends on the initial coordinates $(x_1, z_1, t_1)$ of the atom cloud at the first laser pulse.    Since the atom wavefunction is spatially delocalized, the final phase shift will vary with position across the atom.  These position dependent phase shifts  lead to an atom probability distribution (in one of the output ports) that may be written as $|\Psi(\v{x})|^2=A(\v{x})+B(\v{x})\cos{\left(\phi_0+\phi(\v{x})\right)}$, where $\phi(\v{x})$ is the spatially varying phase shift.  Assuming $|\phi(\v{x})|\ll 1$, the total probability of detecting an atom in this port then depends on the spatial average of the phase over the atom wavefunction:
\be P=\int_{-\infty}^{\infty}|\Psi(\v{x})|^2 dx\approx A + B \cos{\left(\phi_0+\left<\phi(\v{x})\right>\right)} \qquad\qquad |\phi(\v{x})|\ll 1 \ee
where $A \equiv \int_{-\infty}^{\infty}A(\v{x}) dx$, $B \equiv \int_{-\infty}^{\infty}B(\v{x}) dx$, and $\left<\phi(\v{x})\right>\equiv \tfrac{1}{B}\int_{-\infty}^{\infty}B(\v{x})\phi(\v{x}) dx$.

For this noise analysis, we want to calculate the differential response between two atom interferometers separated by a distance $L$ along the laser propagation direction (see Fig. \ref{Fig:satelliteSchematic}).  This gradiometer phase is denoted $\Delta\Phi_t\equiv \left<\Phi_t(x_1=L)\right>-\left<\Phi_t(x_1=0)\right>$, where the expression given in Eq. (\ref{Eqn: Transverse Phase Calc}) has been evaluated at initial positions 0 and $L$ along the $x$ axis and the angle brackets indicate a spatial average over the atom distribution.  Note that in practice the spatially resolved atom detection scheme discussed above (see Sec. \ref{SubSec:Detection}) allows direct resolution of the transverse spatial dependence of the phase shift.  In this section, we assume that we simply average over this information in order to set upper limits on the allowed aberrations of the optical system.

Since we are only concerned with calculating time-dependent phase shifts originating from the jitter amplitudes $\delta z(t)$ and $\delta\theta(k_t, t)$, we may ignore any constant phase shift piece in Eq. (\ref{Eqn: Transverse Phase Calc}) that does not depend on either jitter amplitude.  To enforce this condition, we consider only the first-order jitter susceptibilities $\partial(\Delta\Phi_t)/\partial(\delta z)$ and $\partial(\Delta\Phi_t)/\partial(\delta\theta)$.  We then consider the response at frequency $\omega$ of Eq. (\ref{Eqn: Transverse Phase Calc}) for the position and phase jitter amplitudes, $\delta z(t)  =\widetilde{\delta z}(\omega)e^{- i \omega t}$ and $\delta\theta(k_t, t) =\widetilde{\delta\theta}(k_t, \omega)e^{-i \omega t}$. Assuming that the system can be linearized for small jitter, the transfer function $H_n (\omega)$ that relates each noise source $\delta_n(t)$ to the phase error is simply the first-order susceptibility evaluated at zero jitter amplitude.

With these transfer functions, we can relate the amplitude spectral density of the measured phase $\overline{\delta \phi}$ to the amplitude spectral density of noise sources $\overline{\delta_n}$.  In general, the spectral densities are matrices, but because the measured phase is a scalar quantity and the noise sources are assumed to be uncorrelated, the relationship reduces to
\be
\label{Eq:PSDtoPSD}
\overline{\delta \phi}\,^2 =\sum_{n}|H_n(\omega)|^2 \,\,\overline{\delta_n}\,^2
\ee
where $|H_n(\omega)|$ is the magnitude of the transfer function.  As noted above, the transfer function $H_n (\omega)$ of the particular noise source $\delta_n(t)$ takes the form $\partial(\Delta\Phi_t)/\partial(\delta_n)|_{\{\delta_n\} = 0}$ in this analysis.

By taking the square root of Eq. (\ref{Eq:PSDtoPSD}), we find the amplitude spectral density of the phase noise imprinted on the atom (in $[\text{rad}/\sqrt{\text{Hz}}]$) in terms of the amplitude spectral density of the transverse position jitter $\overline{\delta z}(\omega)$ (in $[\text{m}/\sqrt{\text{Hz}}]$) and of the wavefront variation $\overline{\delta \theta} (k_t,\omega)$ (in $[\text{rad}/\sqrt{\text{Hz}}]$):
\begin{align}
\overline{\delta\phi}(k_t, \omega) & \equiv\sqrt{\left(  \left| \frac{\partial(\Delta\Phi_t)}{\partial(\delta z)} \right| \overline{\delta z}(\omega)  \right)^2 + \left(  \left| \frac{\partial(\Delta\Phi_t)}{\partial(\delta\theta)}\right| \overline{\delta\theta}(k_t, \omega)  \right)^2} \label{Eqn: TransversePhaseNoiseQuadrature}\\
&=2 \, N\sqrt{\left(\tfrac{2\pi}{\lambda_t}\theta(k_t)\overline{\delta z}\right)^2+\overline{\delta\theta}^2}\left(5-9\cos\!{\left[\frac{2 \pi \Delta z}{\lambda_t}\right]}\cos\!{\left[2 \omega T\right]}+4\cos\!{\left[3 \omega T\right]}\right) \sin\!{\left[\frac{\pi L \lambda}{2\lambda_t^2} \right]}\exp\!\left[{-\frac{\pi^2\sigma^2}{4\lambda_t^2 \ln\!{2}}}\right]\label{Eqn: TransversePhaseNoiseResponse}
\end{align}
where the noise contributions from transverse motion and dynamic wavefront variation have been summed in quadrature since they are assumed to be uncorrelated.  The atom position distribution is taken to be a Gaussian with width $\sigma$ (FWHM), $N=k_\text{eff}/2k$ is the order of the LMT beamsplitter, and $T$ is the interrogation time.  The Coriolis-induced transverse position separation of the upper and lower interferometer paths is given by $\Delta z$.  A five-pulse sequence is assumed in obtaining Eq. (\ref{Eqn: TransversePhaseNoiseResponse}).  Finally, Eq. (\ref{Eqn: TransversePhaseNoiseResponse}) represents the magnitude of the response with respect to the aberration phase $\Theta_t$ (i.e., the quadrature sum of both sine and cosine aberrations).

We note that as $\Delta z \rightarrow 0$, the frequency dependence of the wavefront noise given in Eq. (\ref{Eqn: TransversePhaseNoiseResponse}) is identical to the GW phase response for a five-pulse sequence (see Eq. (\ref{Eqn: GW phase})).  This is expected since both effects arise from what amounts to a time variation in the local phase of the laser.  However, as opposed to the GW signal, the response to wavefront aberration is suppressed to the extent that the aberrations are common between the two interferometers.

\subsubsection{Transverse Vibration Induced Phase Noise}

\begin{figure}
\begin{center}
\subfigure[ ]{\label{Fig:TransversePhaseReqDZ}
\includegraphics[width=3.0 in]{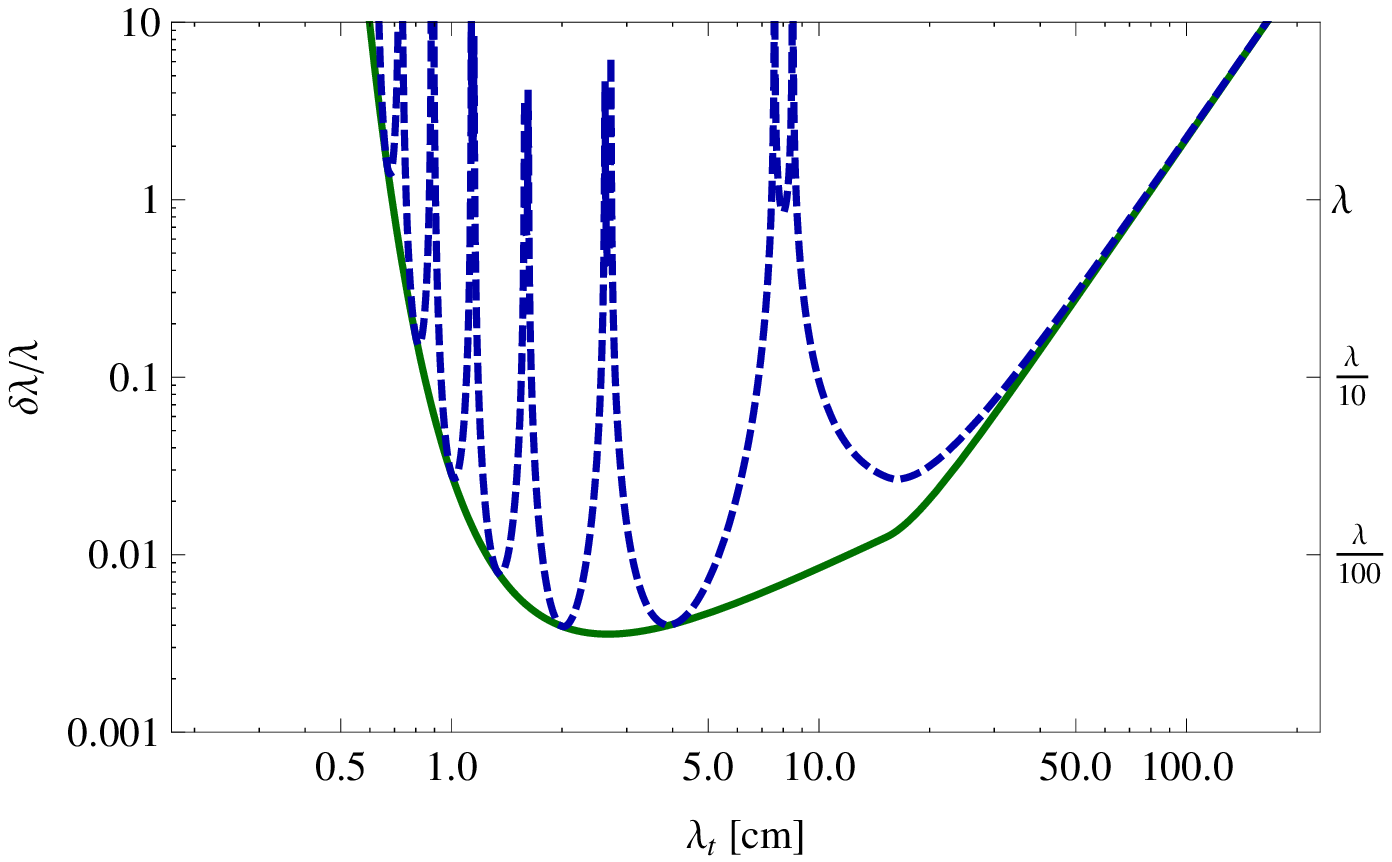}}
\subfigure[ ]{\label{Fig:TransversePhaseReqDPHI}
\includegraphics[width=3.0 in]{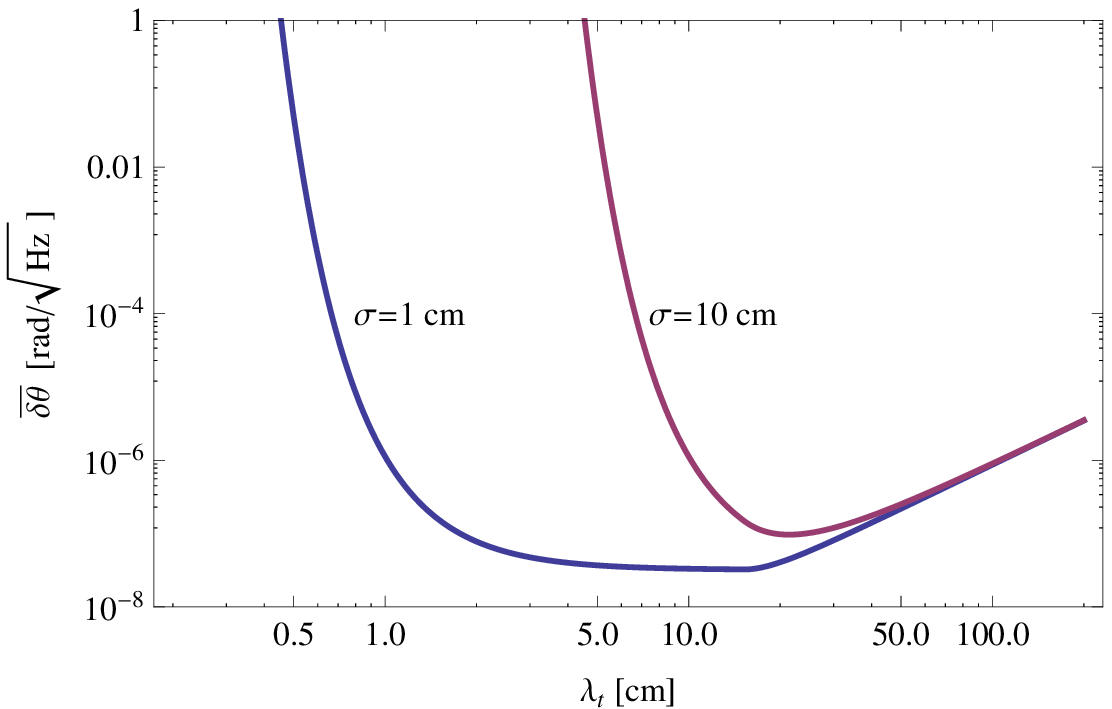}}
\caption{Transverse spatial phase profile requirement as a function of transverse wavelength. Both figures assume the five-pulse interferometer of Fig. \ref{Fig:FivePulseSequence} with $200 \hbar k$ atom optics, $L=30~\text{km}$ baseline, and a $\overline{\delta\phi}=10^{-4}~\text{rad}/\sqrt{\text{Hz}}$ phase noise requirement.  (a) Static wavefront requirements due to satellite transverse motion with amplitude $\overline{\delta z} = 10~\text{nm}/\sqrt{\text{Hz}}$ at frequencies within the GW detection band versus spatial wavelength of the perturbation. The atom cloud size is taken to be $\sigma=1~\text{cm}$ (FWHM). The blue (dashed) curve assumes a leader-follower orbit, while the solid (green) curve assumes an inclined-great-circles orbit. The large notches in the dashed (blue) envelope arise from resonant cancellations of the wavefront aberration with the large transverse separation of the leader-follower interferometer paths. Both curves have been partially enveloped.  (b) Temporal stability requirements for wavefront perturbation amplitude versus spatial wavelength of the perturbation.  This constraint applies only to perturbation amplitude variations that occur at temporal frequencies within the GW detection band.  The pair of curves show the effect of averaging over the atom cloud size. As labeled, the curves assume atom cloud widths of $\sigma=1~\text{cm}$ (FWHM) and $\sigma=10~\text{cm}$ (FWHM). Both curves have been enveloped.  \label{Fig: Earth diff}  }
\end{center}
\end{figure}

Here we focus on the wavefront-induced noise due to stochastic variation of the relative transverse position $\delta z$ of the atom with respect to the aberrated beam during the interferometer pulse sequence.  This type of transverse position jitter may result from variations in the satellite position.  Sensitivity to transverse position jitter of the satellite is given by Eq. (\ref{Eqn: TransversePhaseNoiseResponse}) with $\overline{\delta\theta}(\omega)=0$.  In order for this noise source to be less than the atom shot noise $\overline{\delta\phi}$, the maximum allowable wavefront aberration (in fractions of an optical wave) is given by
\be\frac{\delta\lambda}{\lambda}\equiv\frac{\theta_0(k_t)}{2\pi}=\frac{1}{\pi^2}\frac{\overline{\delta\phi}\,\lambda_t}{N\,\overline{\delta z}}\csc\!{\left[\frac{\pi  L \lambda }{2 \lambda_t^2}\right]}\exp\!{\left[\frac{\pi^2\sigma^2}{4\lambda_t^2 \ln\!{2}}\right]}\left(62+63\cos\left[\frac{2 \pi \Delta z}{\lambda_t}\right]\right)^{-1} \label{Eqn: Wavefront delta z}\ee
assuming $k_\text{eff}=2N k=200 k$ LMT atom optics, $\overline{\delta\phi}=10^{-4}~\text{rad}/\sqrt{\text{Hz}}$, and motion $\delta z$ at the GW peak sensitivity frequency $\omega_c=\tfrac{2}{T} \cos^{-1}{\!\sqrt{\tfrac{3}{8}}}$.  Figure \ref{Fig:TransversePhaseReqDZ} shows the general wavefront constraint for all spatial wavelengths due to $\delta z$ vibration with amplitude $\overline{\delta z}=10~\text{nm}/\sqrt{\text{Hz}}$ for both the leader-follower and inclined-great-circle orbits. Note the expected exponential relaxation of the wavefront constraint due to spatial averaging for $\lambda_t<\sigma$ as well as the expected corner wavelength at $\lambda_t\sim\sqrt{L \lambda}$ above which the perturbations start to be common for the two interferometers.   For wavelengths in the intermediate region $\sigma<\lambda_t<\lambda_c$, which are not suppressed by spatial averaging and are not common between the interferometers, the maximum allowable wavefront aberration in waves is $\delta \lambda \sim \frac{\lambda}{1000} \left(\frac{\lambda_t}{\text{cm}}\right)\left(\frac{10~\text{nm}/\sqrt{\text{Hz}}}{\overline{\delta z}}\right)$.  For the leader-follower orbit, however, the larger transverse position separation ($\Delta z \sim 4 \text{ cm versus } 80 \; \mu\text{m}$) results in resonant cancellation of certain transverse wavelengths. In the vicinity of these particular wavelengths, the wavefront requirements are significantly relaxed.

\begin{table}
\setlength{\tabcolsep}{5pt}
\begin{tabular}{lll}
Parameter & Specification & Location in text\\
\hline
\hline
Transverse Position & $< 10 \;\frac{\text{nm}}{\sqrt{\text{Hz}}}$ & Wavefront Aberration (Sec. \ref{SubSec:WavefrontAberrations})\\
Angle Jitter & $< 1\;\frac{\text{nrad}}{\sqrt{\text{Hz}}}$ & Laser Pointing Angle Jitter (Sec. \ref{subsec:PointingJitter})\\
Angular Rate & $< 1 \text{ nrad/s}$ & Rotational Effects (Sec. \ref{subsection:ErrorModel})\\
\end{tabular}
\caption{Satellite position and angle control requirements. These requirements form constraints on one satellite's laser position and angle relative to the counter-propagating beam from the opposing satellite. We therefore can split the requirements into a coarse constraint on the satellite and a fine constraint on the laser's mounting platform. The constraints apply to both leader-follower and inclined-great-circle orbits.}
\label{tab:SatelliteControl}
\end{table}

\begin{figure}
\begin{center}
\includegraphics[width=4.0 in]{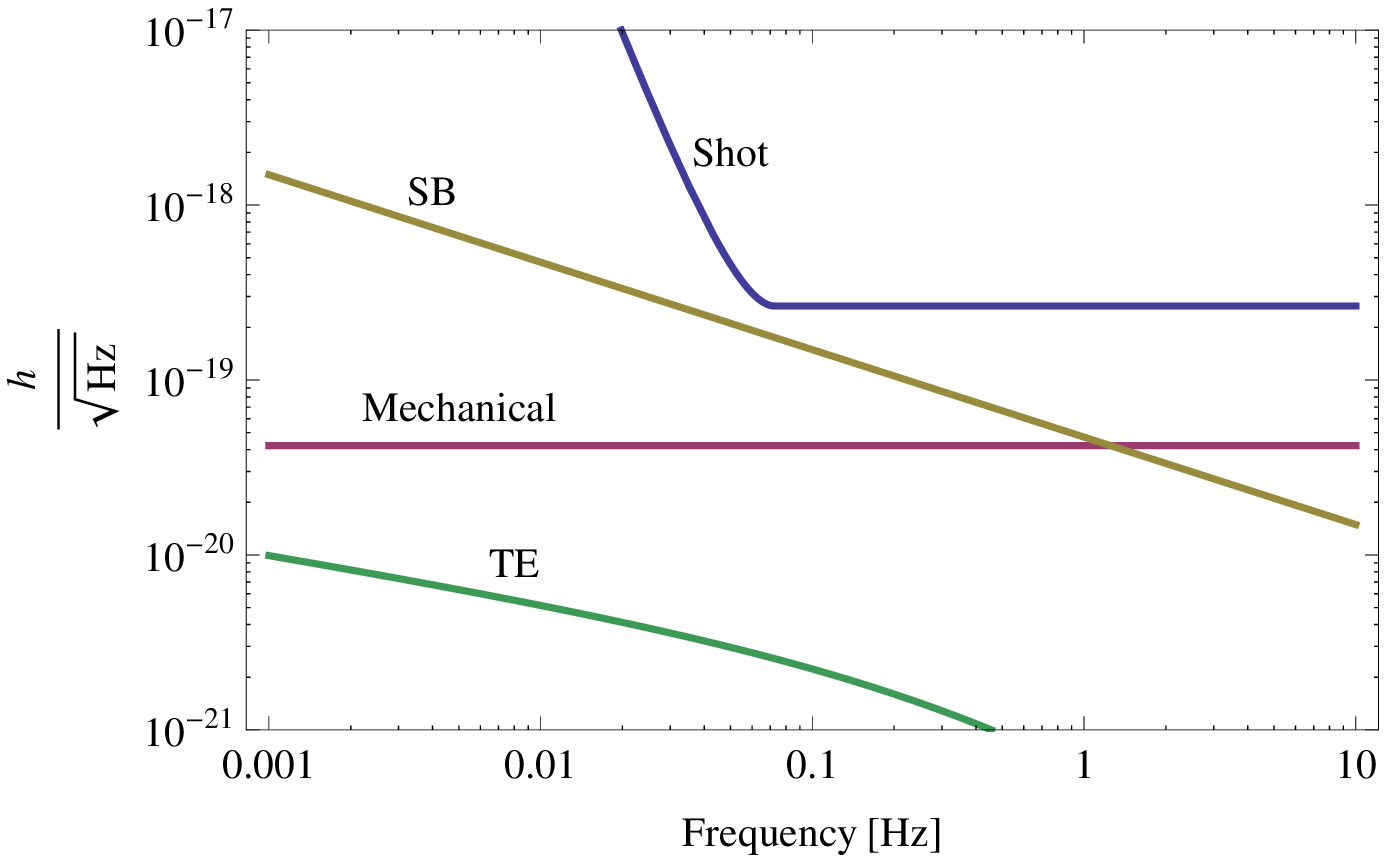}
\caption{ \label{Fig:WavefrontJitterSources} Effective strain induced by several sources of wavefront noise.  Each of the effects are caused by a phase aberration that is imparted to the laser beam as it reflects off the final telescope mirror.  ``Mechanical'' (red) is the off-resonant mechanical acceleration noise of the mirror assuming a normal mode with $\omega_n=2\pi \times 2~\text{kHz}$ and $Q_n=10^2$ and a white noise acceleration amplitude spectral density of $10^{-7}g/\sqrt{\text{Hz}}$.  ``SB'' (yellow) is substrate Brownian noise at temperature $\tau=300~\text{K}$ for a $m=1~\text{kg}$ mirror assuming a normal mode with $\omega_n=2\pi \times 2~\text{kHz}$, $Q_n=10^2$, and with effective mass fraction $\alpha_n=10^{-3}$.  ``TE'' (green) is thermoelastic noise for a beryllium mirror substrate at temperature $\tau=300~\text{K}$ assuming a Gaussian deformation scale factor of $w_0=1~\text{cm}$.}
\end{center}
\end{figure}

\subsubsection{Dynamic Wavefront Induced Phase Noise}

Temporal fluctuations of the spatial mode can also be a problem if the mode varies at frequencies within the GW detection band.  The phase noise induced by such dynamical wavefront variations is given by Eq. (\ref{Eqn: TransversePhaseNoiseResponse}) with $\overline{\delta z}(\omega)=0$.  Figure \ref{Fig:TransversePhaseReqDPHI} shows the required constraint on the fluctuation amplitude $\overline{\delta\theta}(k_t, \omega_c)$ in order for this noise source to be less than the atom shot noise.  To be maximally conservative, the wavefront fluctuation is assumed to be at the GW peak sensitivity frequency $\omega_c$.

To put the dynamical wavefront constraints given in Fig. \ref{Fig:TransversePhaseReqDPHI} into perspective, we consider several well-known sources of wavefront fluctuation associated with aberrations caused by the mirrors in the optics system \cite{numata}.  To be concrete, we model the wavefront aberrations caused by vibrations and thermal fluctuations of the $30~\text{cm}$ primary telescope mirror.  The primary mirror is significant since it is the last element that the beam encounters before propagating to the atoms, and no wavefront spatial filtering can be done after this element.  We emphasize that the same analysis can easily be extended to other elements in the optics system as necessary.

Fluctuations of wavefront aberrations are a source of stochastic noise in the interferometer.  As a result, they contribute to the strain noise floor of the gravitational wave detector.  Figure \ref{Fig:WavefrontJitterSources} shows the effective strain noise contributed by the various sources of dynamical wavefront aberrations that we model.  For comparison, the atom shot-noise limited strain sensitivity from Fig. \ref{Fig:sensitivityAGISLEO} is reproduced in blue (labeled ``Shot'').

The first effect we consider is dynamic wavefront aberrations caused by vibrations of the primary telescope mirror.  We consider a mirror modeled as a solid cylindrical disk with diameter $D=30~\text{cm}$.  The mirror surface is nominally polished so as to produce a collimated Gaussian beam upon reflection (see Fig. \ref{Fig:AGIS-Telescope} for example).  However, the surface of the mirror can become deformed as a result of vibrations of the mirror substrate, and these aberrations will be imprinted onto the laser beam as a spatially dependent phase.  It is convenient to decompose the mirror vibrations into a set of normal modes and consider the effect of each mode separately.  For example, the lowest frequency mode corresponds to the bowl-shaped deformation given by $\delta x(r)=\delta x_0 J_0(2 j_{0,1}r/D)$, where $J_0$ is the zeroth order Bessel function of the first kind, and $j_{0,1}$ is the first zero of $J_0$.  (For simplicity, we assume Dirichlet boundary conditions on the mirror edge.)  In order to leverage the results of Fig. \ref{Fig:TransversePhaseReqDPHI}, this deformation can be rewritten as a discrete Fourier series.  Since the first term in the series has a $97\%$ overlap with the deformation, the mirror surface profile is well-approximated by the truncated series $\delta x(r)\approx \delta x_0 \cos{(\pi r/D)}$.  Therefore, excitations of the the lowest order normal mode of the mirror correspond to dynamic wavefront perturbations with transverse spatial wavelength $\lambda_t=2D\approx 0.6~\text{m}$ in Fig. \ref{Fig:TransversePhaseReqDPHI}.

The normal modes of the mirror can be excited by vibration of the satellite.  We conservatively assume a white noise acceleration spectrum of $\overline{\delta a} =10^{-7}g/\sqrt{\text{Hz}}$ which acts as a driving force for this motion \cite{vallado}.  The vibration response to acceleration of one of the normal modes is a narrow Lorentzian centered at the resonant frequency of the mode:
\be V_{n}(\omega)\equiv\frac{\delta x_0}{\delta a}=\frac{1}{\sqrt{(\omega_n^2-\omega^2)^2+(2\gamma_n \omega_n)^2}}\ee
Here $\gamma_n$ is the (structural) damping of the mode with resonance frequency $\omega_n$.  Even the lowest of the mirror resonances (corresponding to the $\lambda_t=2D$ mode discussed above) tends to be relatively high frequency ($\omega_n>100~\text{Hz}$) and sharply peaked ($Q_n\equiv\tfrac{1}{2\sqrt{3}\gamma_n}>10^2$), so the response of the mirror surface in the GW detection bandwidth is relatively flat, corresponding to the tail of the Lorentzian.  The phase noise induced by these off-resonant mechanical vibrations is given by Eq. (\ref{Eqn: TransversePhaseNoiseResponse}) with $\overline{\delta z}(\omega)=0$ and $\overline{\delta\theta}=k\,V_{n}(\omega) \overline{\delta a}$.  The resulting effective strain noise from mechanical vibrations of the lowest mirror mode (with $\lambda_t=2D$) is shown in red (labeled ``Mechanical'') in Fig. \ref{Fig:WavefrontJitterSources}.

The effect of higher order normal mode oscillations is generally suppressed since these modes have correspondingly higher resonance frequencies, and the size of the mechanical response in the GW band scales as $\sim \omega_n^{-2}$.  However, since they are also associated with smaller $\lambda_t$, these modes can induce a larger intrinsic response (see Fig. \ref{Fig:TransversePhaseReqDPHI}).  Specifically, for long spatial wavelengths $\lambda_t > \lambda_c$, the phase response scales as $\lambda_t^{-2}$.  Assuming a linear dispersion relation for the mirror modes, the transverse wavelength of the mode goes as $\lambda_t\sim \omega_n^{-1}$.  As a result, the first several normal modes make roughly equal contributions to the noise floor.  Beyond this, higher frequency normal modes with transverse wavelengths $\lambda_t<\lambda_c$ are suppressed by the decreased mechanical response and (eventually, for $\lambda_t<\sigma$) by spatial averaging over the atom cloud.

In addition to external mechanical excitation, the normal modes of the mirror also vibrate as a result of the finite temperature of the substrate\cite{numata, raab}.  These thermally excited mirror vibrations are known as substrate Brownian (SB) noise.  The phase noise caused by substrate Brownian noise is given by Eq. (\ref{Eqn: TransversePhaseNoiseResponse}) with $\overline{\delta z}=0$ and $\overline{\delta\theta}=\delta\theta_\text{SB}$ where
\be\delta\theta_\text{SB}=k \,\overline{\delta x}_n\approx k \sqrt{\frac{4 k_B \tau}{m \alpha_n \omega_n^2}}\sqrt{\frac{\gamma_n}{\omega}} \ee
\noindent and where $\overline{\delta x}_n$ is the amplitude spectral density of the vibration amplitude of the mirror mode with resonance frequency $\omega_n$, structural damping $\gamma_n$, and effective mass $m \alpha_n$ \cite{raab}.  This result follows from the fluctuation dissipation theorem for a mirror substrate of temperature $\tau$ \cite{levin}.  Here $m$ is the mass of the mirror itself and $\alpha_n$ quantifies the effective fraction of the mirror mass that oscillates for a particular mode.  Values for the effective mass $m \alpha_n$ of the mode depend on the details of the mode shape as well as material properties of the substrate and the geometry of the mirror.  As an example, the values of $\alpha_n$ reported by \cite{raab} range from $10^{-3}$ to $10$ for a $10~\text{cm}$ diameter, $8.8~\text{cm}$ thick fused silica mirror with mass $m=1.6~\text{kg}$.  For a conservative estimate, we take $m=1~\text{kg}$ and $\alpha=10^{-3}$.  The resulting effective strain noise caused by substrate Brownian noise of the the lowest mirror mode (with $\lambda_t=2D$) is shown in yellow (labeled `SB') in Fig. \ref{Fig:WavefrontJitterSources}.  As with mechanical noise above, the first few higher mirror modes (which have smaller $\lambda_t$) are potentially of comparable size since the phase response scales as $\lambda_t^{-2}$.  However, the overall response for the different modes will depend on the detailed scaling of $\alpha_n$ for the specific mirror geometry, and this is beyond the scope of our analysis.

Thermal fluctuations in the mirror substrate can also cause wavefront aberrations by inducing non-uniform thermal expansion, an effect known as thermoelastic noise \cite{pinard, numata, levin}.  This effect depends strongly on the material properties of the mirror substrate, such as the coefficient of thermal expansion, thermal conductivity, density, and specific heat.  As an example, we consider the thermoelastic noise for a beryllium mirror at $300~\text{K}$.  Note that this is a conservative estimate, since the James Webb Space Telescope (JWST) will use a beryllium mirror at $30~\text{K}$, and the lower temperature significantly reduces thermoelastic noise \cite{edinger}.  The power spectral density of thermoelastic vibrations has previously been calculated for a Gaussian deformation with scale factor $w_0$ \cite{numata, pinard}.  To be maximally conservative, we choose the scale factor to be $w_0=1~\text{cm}$ since this maximizes the thermoelastic amplitude spectral density in the GW frequency band for the beryllium mirror.  Represented in the Fourier domain, a Gaussian surface deformation contains a distribution of transverse wavelength components $\lambda_t\gtrsim w_0$.  The total phase noise response to the Gaussian deformation is then the sum of each spectral component weighted by the amplitude of the phase noise spectral response $\overline{\delta\phi}(k_t, \omega)$ given by Eq. (\ref{Eqn: TransversePhaseNoiseResponse}).  As a conservative simplification, we take the value of the peak response of $\overline{\delta\phi}(k_t, \omega)$ which occurs at $\lambda_t\approx 10~\text{cm}$ as the response for the entire Gaussian spectrum.  The resulting effective strain due to thermoelastic noise is shown shown in green (labeled ``TE'') in Fig. \ref{Fig:WavefrontJitterSources}.

The dynamic wavefront noise effects considered so far are all caused by a phase aberration that is imparted to the laser beam as it reflects off the final telescope mirror.  Alternatively, wavefront noise may be present on the beam as a result of the intrinsic limitation of the spatial mode purity of the laser source.  To eliminate this type of aberration, the laser beam can be sent through a high finesse mode-scrubbing optical cavity prior to delivery to the final telescope optics.  In this case, the wavefront noise upon exiting the optical cavity will be limited by the same sources of dynamic wavefront noise shown in Fig. \ref{Fig:WavefrontJitterSources} as applied to the mirrors in the optical cavity, rather than by anything intrinsic to the laser.

\subsubsection{Atom Distribution Induced Phase Noise}

\label{SubSubSec:ShotToShot}

Quantum projection noise and variation in the preparation of the atom sample result in shot-to-shot fluctuations of the atom cloud size and its position in the interferometer beam. At zero temperature, if the two paths of the interferometer do not separate transversely, the cloud size variation induces no net phase error as both paths average over the same wavefront error at each pulse. Such a constant phase is perfectly cancelled by the pulse sequence. Due to finite temperature, however, the expansion of the cloud will result in a phase error since the cancellation will no longer be perfect between pulses. Moreover, the interferometer paths do separate in general, and each path explores a slightly different region of the wavefront aberration, resulting in non-zero phase shift. If the cloud size or transverse position changes from shot to shot, this phase-shift varies in time, mimicking a gravitational wave.

We compute the phase noise due to stochastic variations in the atom clouds' initial size and position. The size during the interferometer is determined by the linear expansion associated with the clouds' finite temperature. Assuming the typical FWHM cloud size and the corresponding amplitude spectral density for the two distant atom clouds are identical, but uncorrelated, and assuming that the amplitude spectral density of the transverse position is also identical and uncorrelated, we find that the maximum allowable wavefront aberration is proportional to $\csc\!{\left[\pi \Delta z/\lambda_t \right]}^2$. This scaling factor indicates that transverse wavelengths longer than the transverse separation will be quadratically suppressed, as expected.  As previously mentioned, the cloud expands linearly once released from the trap due to its non-zero temperature. At temperatures of $100 \text{ pK}$ and corresponding velocities of $100 \; \mu \text{m/s}$, this effect is negligible.

\begin{figure}
\begin{center}
\includegraphics[width=4.0 in]{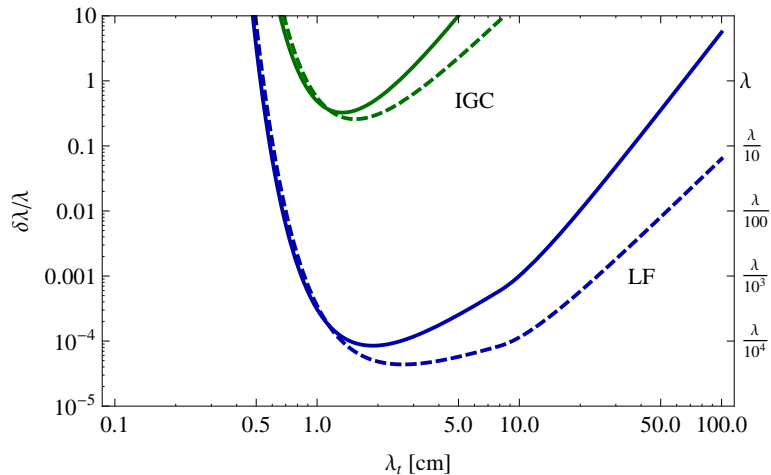}
\caption{ \label{Fig:TransversePhaseShotToShot} Transverse spatial phase profile requirement as a function of transverse wavelength for shot-to-shot noise sources. All curves have been enveloped and assume the five-pulse interferometer of Fig. \ref{Fig:FivePulseSequence} with $200 \hbar k$ atom optics, an $L=30~\text{km}$ baseline, and a $\overline{\delta\phi}=10^{-4}~\text{rad}/\sqrt{\text{Hz}}$ phase noise requirement. The clouds are assumed to be $1 \text{ cm}$ FWHM, and the amplitude spectral densities of the cloud size and cloud centroid position noise are both $1 \; \mu\text{m}/\sqrt{\text{Hz}}$ (This assumes a $f_r=1~\text{Hz}$ repetition rate with $10^8$ atoms per shot). The curves associated with the leader-follower and inclined-great-circles orbits have been labeled ``LF'' (blue) and ``IGC'' (green) respectively. The solid curves are associated with fluctuations in initial atom cloud size, while the dashed curves are associated with fluctuations in initial cloud position in the beam. Note that the wavefront requirements shown here are maximally conservative; those from the LF curves in particular can be significantly mitigated by characterizing the wavefront and by reducing the transverse separation through multiple-pulse interferometer sequences (see Sec. \ref{SubSubSec:WavefrontMitigation}).}
\end{center}
\end{figure}

The transverse trajectory deflection of the inclined-great-circles orbit is suppressed by $\sim L/R \sim 10^3$ compared to that of the leader-follower orbit (see Fig. \ref{Fig:TransverseMotionComparsion}). Consequently, for the same wavefront aberration, the inclined-great-circle orbit has a significantly relaxed cloud size and position constraint, as shown in Fig. \ref{Fig:TransversePhaseShotToShot}.  The size of the transverse wavepacket separation ($\Delta z = 4 \text{ cm}$ for the LF orbit, and $\Delta z = 80 \; \mu \text{m}$ for the IGC orbit) is the only difference between the two curves.  In both cases, the shot-to-shot atom cloud size and centroid position variation are assumed to be at the shot-noise limit.  For an atom cloud of $N_a\sim 10^8$ atoms with size $\sigma=1~\text{cm}$, the centroid transverse position uncertainty due to shot-noise is $\delta z\sim\sqrt{N_a}\sigma\sim 1~\text{$\mu$m}$.  Likewise, the cloud size uncertainty is given by $\delta \sigma\sim\sqrt{N_a}\sigma\sim 1~\text{$\mu$m}$ .  These values represent a lower bound on the centroid position and cloud size variation from shot-to-shot.  Since the results of Fig. \ref{Fig:TransversePhaseShotToShot} assume spatially averaged detection, the curves represent maximally conservative wavefront requirements that can be relaxed through the use of more sophisticated detection schemes (see Sec. \ref{SubSubSec:WavefrontMitigation}).

\subsubsection{Wavefront Aberration Noise Mitigation}
\label{SubSubSec:WavefrontMitigation}

In an alternative interferometer beam geometry, the atom optics laser beams can be made to first propagate between two satellite stations along a path that is displaced from the atoms before being redirected to interact with the atoms.  As a consequence, the first propagation segment would serve as a spatial filter, allowing high frequency wavefront noise to diffract out of the beam.  This arrangement is still sensitive to wavefront aberrations caused by the beam steering optics, but these are potentially more well defined than the mode of the laser source.  If needed, this alternative beam geometry could be used in conjunction with a mode-scrubbing cavity.

Commercially available interferometers can measure wavefront aberration with RMS repeatability $\delta\lambda < \lambda/1000$ \cite{4DTechnologies}. In principle, the aberration can be monitored in real time using such a wavefront sensor.  Furthermore, we can leverage the fact that the detection is spatially resolved (see Sec. \ref{SubSec:Detection}) to perform {\it in situ} measurements of the wavefront.  If necessary, in addition to measuring the spatial phase distribution for each of the five-pulse sequences, we can interleave alternate pulse sequences tailored to be maximally wavefront-sensitive.  The initial transverse velocity and the time between pulses can be varied to gain maximal information about the wavefront.

Spatially resolved detection of the atom cloud can help mitigate the wavefront requirements that result from spatial averaging (Fig. \ref{Fig:TransversePhaseShotToShot}).  In particular, the requirements shown in Fig. \ref{Fig:TransversePhaseShotToShot} for the leader-follower orbit exceed the range of current technologies \cite{4DTechnologies}.  However, these bounds are maximally conservative since with spatially resolved detection we can directly measure the shot-to-shot variations of the cloud size and centroid position in real time.  The additional information provided by spatially resolved detection should allow the wavefront requirements shown in Fig. \ref{Fig:TransversePhaseShotToShot} to be relaxed by an amount that depends on how well the shot-to-shot wavefront characterization can be performed.

Finally, to lessen the impact of shot-to-shot noise sources, it is possible to use interferometer sequences with more pulses than considered here to reduce the transverse wavepacket separation. The quadratic scaling of the phase noise with respect to this separation makes it a useful tool for shot-to-shot phase noise management.

\subsection{Laser Phase Noise}
\label{SubSec:LaserPhaseNoise}

As the laser phase is the reference used to measure the atom motion, noise of the laser phase is imprinted as noise on the atom phase.  We consider two sources of laser phase noise: noise of the laser phase $\delta \phi$ and noise $\delta k$ of the center frequency of the laser. For both, the noise is significantly suppressed by using common lasers.

Despite the use of common lasers, the laser phase $\delta \phi$ is asymmetrically imprinted due to the finite light travel time $\frac{L}{c} \sim 100 \;\mu\text{s}$ between the two interferometers \cite{AGIS}. If the laser pulses are longer than the the light travel time, however, the phase noise at only the beginning and end of the pulses will be non-common; the rest of the pulse overlaps in time and is common. Modeling the LMT pulse as a sequence of a large number of pulses, we note that if each component pulse is longer than the finite light travel time, laser phase noise at frequencies lower than $\sim \frac{\pi c}{L} = 30 \text{ kHz}$ will be suppressed.

\begin{figure}
\begin{center}
\includegraphics[width=4.0 in]{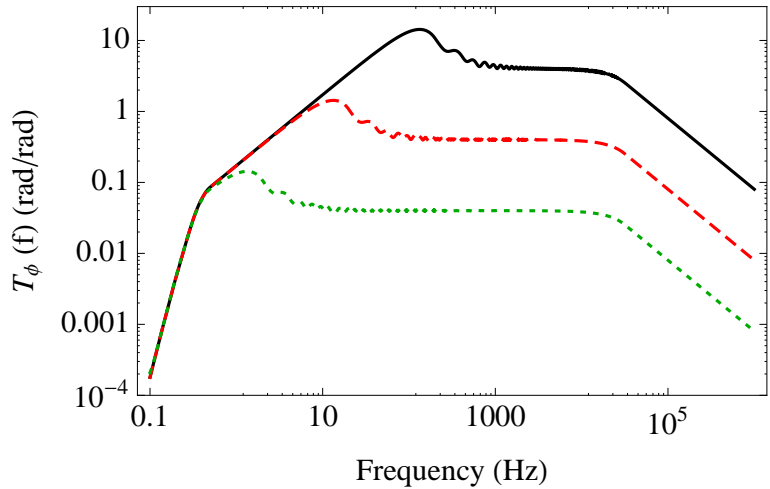}
\caption{ \label{Fig:LaserPhaseNoise} The transfer function relating laser phase noise to atom phase noise as a function of frequency. Sharp oscillations have been enveloped. All three curves have fixed interferometer time $T = 4\text{ s}$ and baseline $L = 30 \text{ km}$. The solid (black), dashed (red), and dotted (green) curves have effective LMT Rabi frequencies of 100, 10, and 1 Hz respectively. Below $\frac{1}{T} =  0.25 \text{ Hz}$, phase noise is suppressed due to the interferometer's insensitivity to static phase offsets and overlapping pulse suppression. The sensitivity to phase noise rises until the atom begins to average the phase above the effective LMT Rabi frequency. This effective low pass filter balances the high pass filter of the overlapping pulse suppression, producing the flat region. Above the corner frequency of the overlapping pulse suppression, $\frac{\pi c}{L} \sim 30 \text{ kHz} $, the atom's averaging suppresses phase noise.}
\end{center}
\end{figure}

In addition, each interferometer is insensitive to static phase offsets and will suppress phase noise at frequencies below $\sim \frac{1}{T}$. Moreover, the atoms average the laser phase at frequencies higher than the effective Rabi frequency of the LMT pulse, $\Omega_\text{st}/N$, where $N$ is the number of pulses. By calculating the quantum evolution of the phase of the atom states through the interferometer sequence, we can analytically find the transfer function $T_{\phi}(f)$ that relates laser phase noise to atom phase noise for a five-pulse sequence (see \cite{Cheinet2005} for a similar calculation).  In Fig. \ref{Fig:LaserPhaseNoise} we show this transfer function for three sample Rabi frequencies $\Omega_\text{st} = 10, 1, 0.1 \text{ kHz} $ for fixed baseline $L = 30 \text{ km}$ and interferometer time $T\sim 4 \text{ s}$. Note the corner frequencies at $0.25 \text{ Hz}$, $\Omega_\text{st}/N$, and $30 \text{ kHz}$ corresponding to the interferometer low-frequency suppression, the atom's LMT Rabi frequency suppression, and the overlapping pulse suppression respectively.

Drift $\delta k$ in the central frequency of the laser between pulses differentially changes the position of the laser phase reference relative to the near and far interferometers. While the near interferometer sees negligible motion of the phase front, the far interferometer sees a wavefront shifted by $\delta k \:L$.  Equivalently, this effect can be understood in the time domain as an extra phase accumulated at a rate $\delta k$ during the time $L/c$ during which the light pulse is not common for the two interferometers.

To place an upper bound on laser phase noise control constraints, consider the maximally conservative Rabi frequency $\sim 10 \text{ kHz }$. As the $\delta \phi$ and $\delta k \:L$ phase noise sources are uncorrelated, they must be added in quadrature as $\sim \sqrt{(\delta \phi)^2 + (\delta k \: L)^2}$.  For this contribution to be $< 10^{-4} \text{ rad/}\sqrt{\text{Hz}}$, the phase noise of the laser must be $< -100 \;\frac{\text{dBc}}{\text{Hz}}$ at an offset of $\sim 20 \text{ kHz }$, and the fractional stability in the laser frequency must be $\sim 10^{-15}(\frac{10^{7} \text{ m}^{-1}}{k})$ over time scales $\sim T$.  Both requirements are experimentally achievable \cite{PhaseNoiseExpt, FreqStabilityExpt}.

Alternatively, in a three or more satellite configuration, two non-parallel beam lines designed to share a common laser can be used to greatly suppress the laser phase noise due to frequency instabilities. Noise from each beam's counter-propagating laser is common to both interferometers on that line, while noise on the shared laser is common to both pairs of interferometers. The gravitational wave signal remains present because it has different components along the two beam lines. The residual phase shift is $\sim \delta k \delta L$, where $\delta L$ is the difference in length of the two beam lines. If the arm lengths are known to $\sim 1 \text{ m}$, then the laser frequency drift may be relaxed to a fractional stability of $\sim 10^{-11}(\frac{10^7 \text{ m}^{-1}}{k})$.

\section{Environmental Considerations in Low Earth Orbit}
\label{sec:EnvCons}

The sensitivity curve in Fig. \ref{Fig:sensitivityAGISLEO} assumes atom shot noise limited phase noise.  Many environmental effects, however, can contribute noise and spurious signals. We analyze these potential backgrounds for both leader-follower and inclined-great-circle orbits.  The error model we consider includes the effects of rotation (Coriolis and centrifugal forces), non-uniform gravity, and magnetic fields.  In addition, the analysis accounts for laser pointing errors and the effect of the finite curvature of the laser wavefronts.

\subsection{Error Model}
\label{subsection:ErrorModel}

To calculate the phase shifts caused by these effects on the atom interferometer, we follow the standard procedure of calculating the contributions from propagation phase, separation phase, and laser phase \cite{varenna, bongs} for both orbital configurations shown in Fig. \ref{Fig:greatCircles}. In both cases, we compute the phase shift between two atom interferometers separated by baseline $L$. The light pulse sequence used for the calculations is the five-pulse, double-diffraction interferometer shown in Fig. \ref{Fig:FivePulseSequence}. The two interferometers are assumed to have an uncontrolled initial velocity difference of $(\delta v_x, \delta v_y, \delta v_z)$ and position offset of $(\delta x, \delta y, \delta z)$, both of which may vary stochastically shot to shot.  Since we are concerned only with terms that vary in time in the detection band, all constant offset phase shift terms have been subtracted.  Here $T_{ij}\equiv-\partial_j g_i$ is the gravity gradient tensor and $g_i$ is the gravitational field vector.  Values for $T_{ij}$ assume a spherical Earth. The results for the leader-follower orbit are presented in Table \ref{tab:TermList} using the coordinate system shown in Fig. \ref{Fig:satelliteSchematic}, where the $x$-axis is the chord connecting the two satellites. The inclined-great-circles configuration results are shown in Table \ref{tab:TermListInclinedGreatCircles}, where the connecting chord is along the $y$-axis as shown in Fig. \ref{Fig:greatCircles}.

Both of these calculations share many commonalities, and we can make the following general statements. Since the atoms are in free fall, many of the largest phase shift terms cancel in pairs.  This occurs because in circular orbit around a spherical Earth, $\Omega_\text{or}^2=g/R= T_\text{xx}=\tfrac{1}{2}T_\text{zz}$, where $R$ is the radius of the orbit and $g$ is the local gravitational acceleration at altitude.  As an example, see terms 1 and 2 in both Table \ref{tab:TermList} and Table \ref{tab:TermListInclinedGreatCircles}.  There are also more elaborate cancellations, such as terms 5, 8 and 9 in Table \ref{tab:TermList}, which together all cancel as a result of $\nabla\cdot\v{g}=-T_{ii}=0$. Additionally, the orbital rotation of the laser axis, $\Omega_\text{or}$ can couple with variations in the atoms' launch velocity and position as well as variations in the laser axis orientation itself to produce many of the dominant backgrounds.

Finally, the $T^4$ scaling of the leading order phase shifts in Table \ref{tab:TermList} is expected given the symmetric triple-loop geometry of the five-pulse sequence (see Fig. \ref{Fig:FivePulseSequence}).  The conventional three-pulse, single-loop ($\tfrac{\pi}{2}$-$\pi$-$\tfrac{\pi}{2}$) gravimeter sequence is sensitive to the quadratic $\sim\tfrac{1}{2}g t^2$ motion of the atom, resulting in a $T^2$ leading-order scaling of the phase shift.  Additional symmetric loops have the effect of making the interferometer sensitive instead to higher-order components of the motion $\sim t^{(\Lambda+1)}$, yielding a $T^{(\Lambda+1)}$ leading phase shift, where $\Lambda$ is the number of loops.  As a consequence, symmetric multiple-loop interferometers respond to higher-order derivatives of the gravitational potential, since the first non-trivial correction due to the derivative $\partial^{(\Lambda)}\phi$ appears as motion proportional to $t^{(\Lambda+1)}$.  A corollary of this is that the response to lower-order derivatives $\partial^{(i)}\phi$ with $i<\Lambda$ is highly suppressed.  Therefore, the use of multiple-loop interferometers can substantially reduce the interferometer's sensitivity to background effects due to inhomogeneous fields, since the perturbations caused by higher-order derivatives are significantly smaller for typical fields.

From the phase shift list for the LF orbit, we note that the largest term that does not cancel (term 12 in Table \ref{tab:TermList}) results from a coupling between gravity gradients and rotation.  This term sets the limit of the size of the transverse velocity jitter $\delta v_z$ for the LF configuration: for $200 \hbar k$ beamsplitters, $4$ s interferometer time, and $\Omega_\text{or} \sim 10^{-3} \text{ rad/s}$, $\delta v_z \lesssim 10^{-8} \text{ m/s}$. This shot-to-shot velocity variation is equivalent to $\lesssim 100 \text{ pK}$ in atom cloud temperature for $10^8$ atoms. Term 13 constrains the laser pointing error rotation rate to $\delta \Omega \sim 1 \text{ nrad/s}$. Term 14 limits the longitudinal position variation of the atoms to $\delta x \sim 10 \;\mu \text{m}$. Term 15 is reduced below the shot noise limit by this latter constraint. These requirements are not strict, but imply a trade-off with several other parameters. If necessary, a small reduction in interferometer time at the expense of bandwidth, for example, could significantly relax these constraints, as both terms scale with multiple orders of the time.

The IGC orbit offers many systematic benefits over the LF orbit. Note that the largest non-cancelling term in the IGC list (term 3 in Table \ref{tab:TermListInclinedGreatCircles}) is the leading term from Table \ref{tab:TermList} but suppressed by an additional power of $\Omega_\text{or} T$. This suppression arises because the IGC interferometer axis is nearly parallel with the orbital rotation axis. Therefore the transverse trajectory deviations due to the Coriolis force are suppressed by $\tfrac{L}{R}$ as compared to the LF orbit, which leads to a phase shift suppression $\propto \Omega_\text{or} T$. For the same stochastic atom position jitter, the contributions to the IGC phase shifts are smaller than those of the LF. The largest phase shift due to atom jitter for the IGC is term 5 in Table \ref{tab:TermListInclinedGreatCircles}, which remains below atom shot noise for position fluctuations of $\delta y \sim 20~\mu \text{m}$ instead of $\sim 10~\mu \text{m}$. Similarly, velocity variations of $\sim 1~\mu \text{m}/\text{s}$ instead of $\sim 10~ \text{nm}/\text{s}$ could be tolerated. However, if the tight atom kinematic constraints can be achieved and $\delta \Omega$ effects can be controlled, then the interrogation time could be doubled to bring these systematics to the same size as atom shot noise.  This would lower the instrument corner frequency by a factor of two, potentially allowing for the detection of lower frequency GWs.

The current IGC analysis does not account for the relative velocity between the two atom clouds, or the required orbital phase offset between the satellites to avoid collision. Both of these effects can be removed by rotating and chirping the interferometer laser to make it remain inertial during the interferometer sequence \cite{varenna, AchimMetrologia}. As mentioned in Section \ref{Sec:SatelliteConfiguration}, the relative velocity between the clouds is useful for Doppler shift frequency tagging of the individual clouds.

It is important to note that the above calculation assumes that the laser beams are plane waves, ignoring the finite curvature of the laser wavefronts.  This is a rough approximation, since the interferometer beams we consider have a Rayleigh range comparable to the satellite separation $L$, and so they are maximally curved for their size at the position of the interferometers.  These effects of wavefront curvature are a correction to the results of Table \ref{tab:TermList}.  The effect of Gaussian curvature is analyzed for high frequency laser pointing jitter in Section \ref{subsec:PointingJitter}, and in this case, the physical beam results in less-stringent pointing requirements than for the plane wave case. The full Gaussian analysis for the above error model would conceivably result in a similar reduction in rotation rate error $\delta\Omega$ sensitivity.

\begin{table}
\setlength{\tabcolsep}{5pt}
\begin{tabular}{lcl}
& Differential phase shift & Size (rad)\\
\hline
\hline
$
 1  $ & $  -60           k_{\text{eff}} L                    \Omega_\text{or}^3            \delta\Omega \, T^4 $ & $ -1.15               $ \\ $
 2  $ & $   60           k_{\text{eff}} L \, T_{\text{xx}}   \Omega_\text{or}              \delta\Omega \, T^4 $ & $ +1.15               $ \\ $
 3  $ & $  888           k_{\text{eff}} L                    \Omega_\text{or}^5            \delta\Omega \, T^6 $ & $ +3.67\times 10^{-4} $ \\ $
 4  $ & $  444           k_{\text{eff}} L \, T_{\text{zz}}   \Omega_\text{or}^3            \delta\Omega \, T^6 $ & $ -3.67\times 10^{-4} $ \\ $
 5  $ & $ -444           k_{\text{eff}} L \, T_{\text{xx}} T_{\text{zz}} \Omega_\text{or}  \delta\Omega \, T^6 $ & $ +3.67\times 10^{-4} $ \\ $
 6  $ & $   30           k_{\text{eff}}                      \Omega_\text{or}^3 \delta v_z                 T^4 $ & $ +1.92\times 10^{-4} $ \\ $
 7  $ & $   15           k_{\text{eff}}      T_{\text{zz}}   \Omega_\text{or}   \delta v_z                 T^4 $ & $ -1.92\times 10^{-4} $ \\ $
 8  $ & $ -444           k_{\text{eff}} L \, T_{\text{xx}}   \Omega_\text{or}^3            \delta\Omega \, T^6 $ & $ -1.84\times 10^{-4} $ \\ $
 9  $ & $ -444           k_{\text{eff}} L \, T_{\text{xx}}^2 \Omega_\text{or}              \delta\Omega \, T^6 $ & $ -1.84\times 10^{-4} $ \\ $
 10 $ & $ -225           k_{\text{eff}} L \, T_{\text{xz}}   \Omega_\text{or}^2            \delta\Omega \, T^5 $ & $ -1.24\times 10^{-4} $ \\ $
 11 $ & $ -\frac{225}{2} k_{\text{eff}} L \, T_{\text{xz}} T_{\text{zz}}                   \delta\Omega \, T^5 $ & $ +1.24\times 10^{-4} $ \\ $
 12 $ & $   15           k_{\text{eff}}      T_{\text{xx}}   \Omega_\text{or}   \delta v_z                 T^4 $ & $ +9.62\times 10^{-5} $ \\ $
 13 $ & $ -\frac{225}{2} k_{\text{eff}} L \, T_{\text{xz}} T_{\text{xx}}                   \delta\Omega \, T^5 $ & $ -6.19\times 10^{-5} $ \\ $
 14 $ & $ -\frac{45}{2}  k_{\text{eff}}                      \Omega_\text{or}^4 \delta x                   T^4 $ & $ -1.67\times 10^{-5} $ \\ $
 15 $ & $   15           k_{\text{eff}} T_{\text{xx}}        \Omega_\text{or}^2 \delta x                   T^4 $ & $ +1.11\times 10^{-5}
$
\end{tabular}
\caption{Differential phase shift error budget for AGIS-LEO in a leader-follower orbit.  The results are based on the triple-loop interferometer in Fig. \ref{Fig:FivePulseSequence}, and the phase difference assumes the gradiometer configuration shown in Fig. \ref{Fig:satelliteSchematic} between two atom interferometers separated by $L=30~\text{km}$ in a leader-follower orbit (see Fig. \ref{Fig:greatCircles}).  An unimportant constant phase shift has been subtracted.  The jitter in the atom initial position and velocity are taken to be $\delta x=1~\mu\text{m}$ and $\delta v_z=10~\text{nm/s}$, respectively.  The laser pointing error rotation rate is $\delta\Omega=1~\text{nrad/s}$.  The orbital radius $R$ and rotational rate $\Omega_\text{or}$ are for a $1000~\text{km}$ orbital altitude.  Interrogation time is $T=4~\text{s}$ and the LMT beamsplitters have $\hbar k_{\text{eff}}=200\hbar k$.
}
\label{tab:TermList}
\end{table}

\begin{table}
\setlength{\tabcolsep}{5pt}
\begin{tabular}{lcl}
& Differential phase shift & Size (rad)\\
\hline
\hline
$
 1  $ & $ -\frac{225}{2} k_{\text{eff}} \frac{L^2}{R} T_{\text{zz}} \Omega_\text{or}^2  \delta\Omega   \, T^5 $ & $ +8.25\times 10^{-5} $ \\ $
 2  $ & $ -\frac{225}{4} k_{\text{eff}} \frac{L^2}{R} T_{\text{zz}}^2            \delta\Omega   \, T^5 $ & $ -8.25\times 10^{-5} $ \\ $
 3  $ & $  \frac{675}{4} k_{\text{eff}} \frac{L^2}{R} \Omega_\text{or}^4                \delta\Omega   \, T^5 $ & $ +6.19\times 10^{-5} $ \\ $
 4  $ & $  \frac{225}{4} k_{\text{eff}} \frac{L^2}{R} T_{\text{yy}}^2            \delta\Omega   \, T^5 $ & $ +2.06\times 10^{-5} $ \\ $
 5  $ & $  \frac{15}{2}  k_{\text{eff}}               T_{\text{yy}}^2 \delta y                  \, T^4 $ & $ +5.57\times 10^{-6} $ \\ $
 6  $ & $   45  \,       k_{\text{eff}} L             T_{\text{yy}}              \delta\Omega^2 \, T^4 $ & $ +7.47\times 10^{-7} $ \\ $
 7  $ & $  \frac{45}{2}  k_{\text{eff}}               T_{\text{yy}}^2 \delta v_y                \, T^5 $ & $ +6.69\times 10^{-7} $ \\ $
 8  $ & $  15            k_{\text{eff}} \frac{L}{R}                   \delta v_x \Omega_\text{or}^3        T^4 $ & $ +3.95\times 10^{-7} $ \\ $
 9  $ & $  \frac{15}{2}  k_{\text{eff}} \frac{L}{R}   T_{\text{zz}}   \delta v_x \Omega_\text{or}          T^4 $ & $ -3.95\times 10^{-7} $ \\ $
 10 $ & $  \frac{15}{2}  k_{\text{eff}} \frac{L}{R}   T_{\text{xx}}   \delta v_x \Omega_\text{or}          T^4 $ & $ +1.98\times 10^{-7}
$
\end{tabular}
\caption{Differential phase shift error budget for AGIS-LEO in an inclined-great-circle orbit.  The results are based on the triple-loop interferometer in Fig. \ref{Fig:FivePulseSequence}, and the phase difference assumes the gradiometer configuration shown in Fig. \ref{Fig:satelliteSchematic} between two atom interferometers separated by maximum distance $L=30~\text{km}$ in an inclined-great-circles orbit (see Fig. \ref{Fig:greatCircles}). The required $\sim 100~\text{m}$ offset to avoid collision is not included.  An unimportant constant phase shift has been subtracted.  The jitter in the atom initial position and velocity are taken to be $\delta y=1~\mu\text{m}$ and $\delta v_x=\delta v_y=10~\text{nm/s}$, respectively.  The separation distance $L$ is taken to be $30~\text{km}$, which is the maximum separation over the course of the orbit, and the point of largest systematic error. The laser pointing error rotation rate is $\delta\Omega=1~\text{nrad/s}$.  The orbital radius $R$ and rotational rate $\Omega_\text{or}$ are for a $1000~\text{km}$ orbital altitude.  Interrogation time is $T=4~\text{s}$ and the LMT beamsplitters have $\hbar k_{\text{eff}}=200\hbar k$.
}
\label{tab:TermListInclinedGreatCircles}
\end{table}

\subsection{Orbital Altitude}
\label{sec:orbAlt}

The error model in the previous section indicates that many systematic effects are mitigated at larger orbital radii, where the rotation rate, $\Omega_\text{or}$, is smaller.  In general, the selection of the satellites' orbital altitude needs to account for several factors, including vacuum requirements, magnetic fields, gravity gradients, rotational effects, and the fraction of an orbital period spent in the Earth's shadow.  Of these, only the last favors lower orbits, since without a satellite sunshield we rely on the Earth's umbra to protect the interferometers' atom clouds from a large $\sim 10 \; \text{s}^{-1}$ scattering rate with solar photons \cite{AGIS}.  For an equatorial orbit at $1000~\text{km}$, a satellite spends approximately one third of its time in the Earth's shadow.  All of the other altitude-influencing factors mentioned improve with increasing height, and in the following subsections we argue that their effects are acceptable at or even below $1000~\text{km}$.

\subsection{Vacuum Requirements}
Collisions with particles in the background gas cause the decoherence of atoms in the interferometer region.  While this is not a source of noise and does not introduce a spurious phase shift, the removal of atoms from the cloud reduces the sensitivity of the instrument.  Around $700~\text{km}$ above the Earth's surface, the atmosphere is dominated by H, He, and atomic O with densities of $10^6~ \frac{\text{atoms}}{\text{cm}^3}$ at temperatures near 1000 K \cite{MSISE90}.  This corresponds to pressures of $10^{-10}$ Torr, typical for low-background-collision-loss atom interferometry experiments.  At this pressure, the average time between collisions of an atom with the background gas is $\sim 100 \text{ s}$ \cite{AGIS}, several times longer than the proposed interferometer time of $\sim 10 \text{ s}$.  Thus, satellites constructed with low-outgassing materials at altitudes above $700~\text{km}$ are sufficient to meet the vacuum requirements of AGIS-LEO.

\subsection{Magnetic Fields}
Since the atom interferometers take place outside of the AGIS-LEO satellites, the atoms are in a region of space with uncontrolled magnetic fields. The local magnetic field defines the quantization axis used for the atom-laser interaction and shifts the atomic energy levels, which can cause an unwanted phase shift in the atom interferometer. Variations in the direction of the magnetic field have been addressed in the original AGIS proposal \cite{AGIS}, and similar strategies to define the quantization axis, such as a large permanent magnet, can be employed in AGIS-LEO.

Phase shifts due to time-dependent variation in the atomic energy levels are more concerning. In order to mitigate these effects, we perform interferometry on atoms in the $\left|m=0\right>$ clock state which is first-order insensitive to energy shifts from magnetic fields. However, alkali atoms have a second order energy shift $\Delta E = \frac{1}{2} \alpha \vec{B}^2$, and since $\vec{B}$ varies between the two AGIS-LEO interferometers, there is a differential phase shift between them. We analyzed the in-situ magnetic field for various AGIS-LEO orbits using the World Magnetic Model 2010 (WMM2010) \cite{WMM2010}. The WMM2010 models the Earth's magnetic field using a 12th order spherical harmonic expansion with the minimum wavelength of $28.8^{\circ}$. This spatial resolution is much larger than the orbital arc traversed during a $24~\text{second}$ interferometer sequence, so we can model the field as $\vec{B_0} + \frac{\partial \vec{B}}{\partial \lambda} \lambda(t-t_0)$ where $\lambda$ is the orbital longitude. The leading order response of a single three-pulse atom interferometer to these spatial gradients is:

\be \Delta\phi_\text{B} \simeq \frac{\hbar k_\text{eff}}{m} \alpha T^2 B_0 \frac{\partial B}{\partial \lambda} \frac{\partial \lambda}{\partial x} \simeq \frac{\hbar k_\text{eff}}{m} \alpha T^2 B_0 \frac{\partial B}{\partial \lambda} \frac{1}{R_E}\label{Eqn: mag phase}\ee

For this long-wavelength model, variations are suppressed by the ratio of the interferometer size to the radius of the Earth, $R_E$. The differential phase shift between two interferometers spaced by a baseline $L$ away from each other is

\be \Delta\phi_\text{B,gradiometer} \simeq \frac{\hbar k_\text{eff}}{m} \alpha T^2 \frac{(\partial_{\lambda} B)^2 L}{R_E^2}\label{Eqn: mag_differential phase}\ee

This is further suppressed by the ratio of the baseline $L$ to the radius of the Earth. For an example $850~\text{km}$ equatorial orbit, using WMM2010, we expect $\sim 200~\text{mG}$ magnetic fields with $\frac{\partial B}{\partial \lambda} \sim 1~\text{mG} / \text{deg. longitude}$. This results in a differential phase shift of $\Delta\phi_\text{B,gradiometer} \sim 10^{-8} \text{ rad}$ which is well below threshold.  Magnetic field dependent phase shifts are expected to be further suppressed by the proposed five-pulse sequence.

Additionally, there may exist finer local variation in the magnetic field due to the presence of the spacecraft. However, the interferometer sequence takes place a distance $d \gtrsim 3~\text{m}$ from the spacecraft, and local spurious magnetic fields fall off as $\sim d^3$. Therefore the magnetic field gradient differences between the two spacecraft can be controlled below our required threshold of $10^{-8}\text{G}/\text{m}$. Furthermore, pulse sequences that use larger numbers of pulses, such as the four- and five-pulse sequences discussed above, will have additional suppressions of this effect.

\subsection{Gravity Gradients}
Fluctuations in the Newtonian gravitational field can masquerade as a gravitational wave. Therefore we have engineered a pulse sequence and use a differential measurement satellite configuration to minimize the phase shift caused by the Newtonian gravitation potential $\phi$. As discussed in Sec. IV A, AGIS-LEO benefits from many exact cancellations due to its pulse sequence symmetry and the orbital symmetry which demands $\Omega_\text{or}^2 = g/R = T_\text{xx} = \frac{1}{2}T_\text{zz}$. The above analysis assumes a spherical earth and does not take into account the local presence of the satellites. We now analyze the impact of a large local mass with position jitter and the effect of the AGIS-LEO constellation orbiting in the Earth's inhomogeneous gravitational field.

\subsubsection{Satellite Position Jitter}
The motion of the satellite is typically a major source of concern for a space-based precision test mass experiment, as it causes large time-dependent gravity gradients that can be a significant source of error. We take advantage of the disposability of the atomic test masses and perform our precision measurement outside of the satellite to decrease our sensitivity to these spurious gravity gradients. Since we are searching for time-dependent gravitational wave signals, we are concerned only with background noise in our detection band; overall static phase offsets between the two atom interferometers are not important. Therefore we require the satellite jitter position $\delta x_\text{sat}(\omega)$ to be small enough in our measurement frequency band so that spurious phase shifts are below $10^{-4}~\text{rad}/\sqrt{\text{Hz}}$.

\begin{figure}
\includegraphics[width=5.0 in]{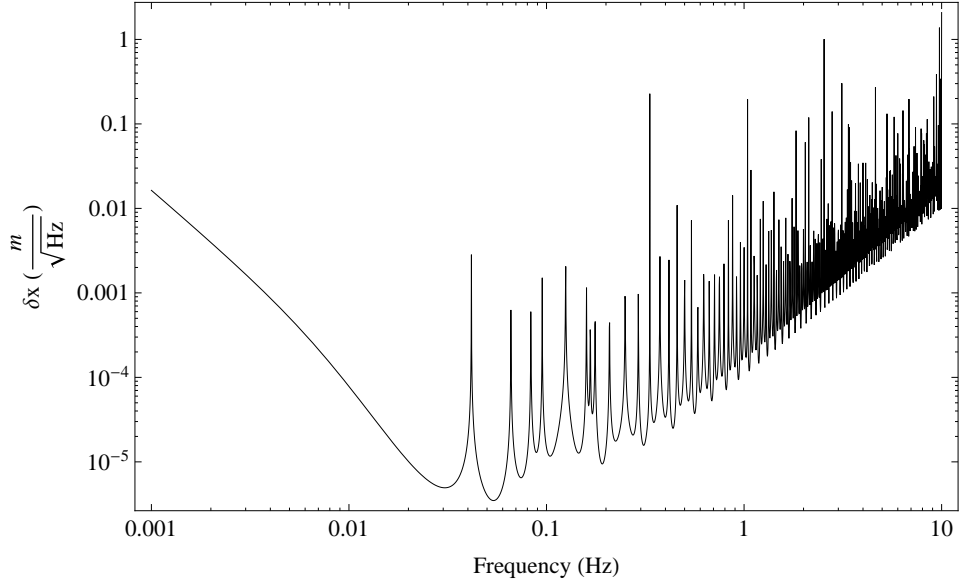}
\caption{ \label{Fig:satelliteJitter} Satellite position jitter requirements as a function of frequency. This plot is for a single interferometer and assumes the five-pulse sequence of Fig. \ref{Fig:FivePulseSequence} with $200 \hbar k$ atom optics, a satellite mass $M = 1000~\text{kg}$, an initial atom cloud distance $d = 10~\text{m}$ from the satellite, and a $\delta\phi=10^{-4}~\text{rad}/\sqrt{\text{Hz}}$ phase noise requirement.}
\end{figure}

We calculate the response of a single atom interferometer to satellite position jitter using perturbation theory \cite{AchimMetrologia}. To first order in the jitter amplitude $\delta x_\text{sat}(\omega)$ the perturbing Lagrangian $\Delta L_\text{sat}$ is

\be \Delta L_\text{sat}(\omega) \simeq \frac{G M_\text{sat} m}{d^2} \delta x_\text{sat}(\omega) \label{Eqn: ggSatelliteJitter}\ee

\noindent where $M_\text{sat}$ is the mass of the satellite, $m$ is the atom mass, and $d$ is the distance between the atom and the satellite. The resulting propagation phase shift \cite{varenna} from this perturbation as a function of frequency determines the interferometer's sensitivity to satellite position jitter. The required satellite control to reduce the position jitter phase shift below atom shot noise is shown in Fig. \ref{Fig:satelliteJitter}. The peak control requirement is $\sim 1~\mu\text{m}/\sqrt{\text{Hz}}$ at $\omega \sim 1/T$ where $T$ is the interrogation time. This requirement is significantly less stringent than that imposed by wavefront aberrations and thus is not a driving factor in the mission design.

\subsubsection{Non-Spherical Earth Gravity Gradients}
The proximity of the Earth's non-uniform gravitational field to the observatory is also a source of non-trivial Newtonian gravitational backgrounds. The near-Earth gravity field is well known and is characterized in the Earth Gravity Model 2008 (EGM2008) by a spherical harmonic expansion \cite{EGM2008}. EGM2008 resolves spatial wavelengths as small as $5~\text{arcminutes}$, which is on the same order as the orbital arc subtended by AGIS-LEO. If the two interferometers are in different gravitational fields, they will record different phase shifts.  These phase shifts will change over the course of the orbit, producing a time-dependent background signal.

As the non-spherical Earth produces a time-varying gravitational potential $\phi$ which can be characterized by spherical harmonics, these spherical harmonics form a natural basis to use for computing the AGIS-LEO response to gravitational deviations from the static, spherical earth. Expanded in this basis, the gravitational potential $\Phi$ is

\be \Phi(r, \theta, \lambda, t) = \frac{G M_E}{r(t)} + \phi(r, \theta, \lambda, t) \ee
\be \phi(r, \theta, \lambda, t) = \frac{G M_E}{r(t)}\left(\sum_{n=2}^{n_\text{max}} \left(\frac{R_E}{r(t)}\right)^n \sum_{m=0}^{n} \left(\bar{c}[n,m]\cos m\lambda(t) + \bar{s}[n,m]\sin m\lambda(t) \right) \bar{P}^n_m[\cos \theta(t)]\right) \label{Eqn: ggSphericalHarmonicExpansion}\ee

\noindent where $M_E$ is the mass of the Earth, $\phi$ is the aspherical potential, $r(t)$, $\theta(t)$, and $\lambda(t)$ are the radial distance, northern spherical polar distance (co-latitude), and longitude of the atom cloud, $\bar{c}[n,m]$ and $\bar{s}[n,m]$ are the coefficients of the EGM2008 model, and $\bar{P}^n_m[\cos \theta(t)]$ are the fully normalized associated Legendre functions of the first kind \cite{EGM2008}. The effects of gravity gradients from a spherical Earth are accounted for in the phase shift calculation presented in Sec. \ref{subsection:ErrorModel}. Therefore we compute the atom interferometer response to an arbitrary spherical harmonic component of $\phi$ using perturbation theory. We compute the response to the following perturbing Lagrangian for the AGIS-LEO leader-follower equatorial orbit

\be \Delta L[n,m,t] = \frac{R_E^n}{r(t)^{n+1}} \left( \bar{c}[n,m]\cos m\lambda(t) + \bar{s}[n,m]\sin m\lambda(t) \right) \ee

We then sum the response of each individual spherical harmonic to produce the total phase shift $\Delta \phi$.

\be \Delta \phi = - \frac{m}{\hbar} G M_E \sum_{n=2}^{n_\text{max}} \sum_{m=0}^{n} \left( \int_{\text{upper}} \Delta L[n,m,t]dt - \int_{\text{lower}} \Delta L[n,m,t]dt \right) \bar{P}^n_m[0] \label{Eqn: ggArbitrarySphericalHarmonic}\ee

\begin{figure}
\includegraphics[width=5.0 in]{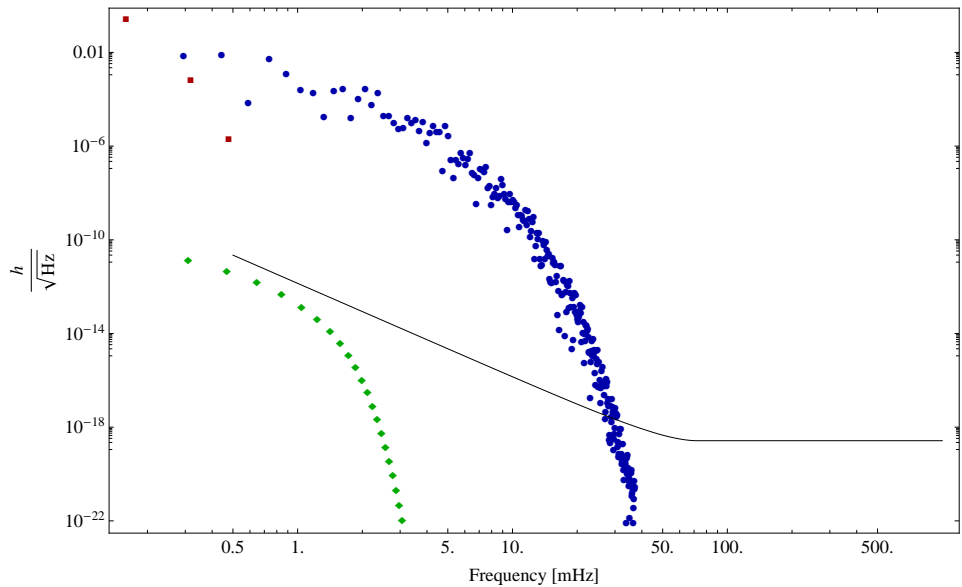}
\caption{ \label{Fig:egm2008SphericalHarmonicResponse} Newtonian Gravity Backgrounds. Each point on this plot is the effective $h/\sqrt{\text{Hz}}$ for a Newtonian gravity background frequency component with a measurement bandwidth of $1~\text{year}^{-1}$. The circles (blue) are EGM2008 components, the squares (red) are phase shifts due to lunar tidal forces, and the diamonds (green) are for free earth oscillations normalized to a $s_{0,2,0}$ peak amplitude of $1~\text{cm}$. This plot is for two interferometers separated by $30~\text{km}$ and assumes the five-pulse sequence of Fig. \ref{Fig:FivePulseSequence} with $200 \hbar k$ atom optics, a $T=4 \; \text{s}$ interrogation time, and a $\delta\phi=10^{-4}~\text{rad}/\sqrt{\text{Hz}}$ phase sensitivity for the AGIS-LEO sensitivity curve.}
\end{figure}

The $105 ~\text{minute}$ orbital period of the AGIS-LEO constellation causes spherical harmonics of order $m$ to induce time-dependent phase shifts with frequency $\omega_m = m(\Omega_\text{or} - \Omega_E)$, where $\Omega_E$ is the Earth's rotation rate. In order to find the full frequency response due to Earth gravity for a $1~\text{year}$ AGIS-LEO science run, we sum the phase shift response for all spherical harmonic components with effective frequency $\omega_m$, and divide by the AGIS-LEO frequency response for a $1~\text{year}^{-1}$ bandwidth. This procedure is identical to that used for plotting gravity wave sources in Figs. \ref{Fig:sensitivityAGISLEO} and \ref{Fig:sensitivity2}. The effective strain response is plotted with circles (blue) on Fig. \ref{Fig:egm2008SphericalHarmonicResponse}. At low frequencies ($f < 30~\text{mHz}$), this Newtonian gravity background is the dominant systematic.

\subsubsection{Lunar and Solar Tidal Forces}
The tidal forces from the Moon and the Sun form a non-negligible contribution to the AGIS-LEO Newtonian gravity background. We treat both sources as perfect spheres and thus their tidal forces arise from gravity gradients, since AGIS-LEO is in free-fall with respect to both the Sun and the Moon and does not see a direct gravitational force from either. While the five-pulse interferometer sequence is insensitive to gravity gradient phase shifts for uniform gradients (i.e. those of the Earth, which it is orbiting), it is sensitive to gradient differences between the satellites. As the satellite constellation orbits the Earth, the orientation of the sensors changes with respect to the vector pointing toward the Moon (or Sun). Therefore the orientation of the Moon gravity gradients $M_\text{ij}$ changes with respect to the interferometer axes, and the two satellite orientations are slightly different. Similarly, one satellite is often farther from the moon than the other, and this creates a difference in size between the gradients experienced by the two sensors, even though their orientations may be the same. Thus $M_{\text{ij},1}$ is not the same as $M_{\text{ij},2}$, which induces a phase shift that is modulated by both the AGIS-LEO and Moon orbital velocities. We compute the phase shift response for a $\delta M_\text{ij} = M_{\text{ij},1} - M_{\text{ij},2}$ using perturbation theory, and the $1~\text{year}$ science run results are shown as squares (red) on Fig. \ref{Fig:egm2008SphericalHarmonicResponse}. The frequency of the response is determined to first order by $\Omega_\text{or}$ and the spatial frequency of the phase response. The Sun produces a similar response, but smaller by $\frac{M_\text{Sun}}{M_\text{Moon}}\frac{D_\text{Moon}^3}{D_\text{Sun}^3}\sim 45\%$.

\subsubsection{Free Earth Oscillations}
The gravitational potential of the Earth is not constant in time and undergoes free earth oscillations at frequencies which are potentially interesting when searching for gravitational waves \cite{JeffreyPark05202005}. Free earth oscillations are spherical harmonic eigenmodes of the Earth which ring at frequencies $< 10~\text{mHz}$ and can be excited by major earthquakes, which can produce $>1~\text{cm}$ height deviations at all points on the Earth's surface \cite{JeffreyPark05202005}. Not only are these signals naturally time-dependent, the orbital motion of the satellite also splits their frequency spectrum, making spectral data analysis techniques more difficult. It is important to note that free earth oscillations are characterized by the geophysics community and will be well-known during AGIS-LEO science runs.

The standard formalism for computing the shape and oscillation frequency for the normal modes of the Earth is well documented in the geophysics community, and we will lean heavily on their analysis \cite{DahlenGlobalSeismology}. The Earth can oscillate in three fundamental classes of modes: radial, toroidal, and spheroidal. Radial modes are characterized by expansion and contraction of the Earth in a purely radial direction with $\v{s} = f(r)\hat{r}$, where $\v{s}$ is the displacement of the Earth's surface. Toroidal modes produce shearing motion between sections of the Earth's surface and have angular variation, but do not contain radial displacement. For a spherical Earth, radial and toroidal modes do not produce a Newtonian gravitational signal for AGIS-LEO. On the other hand, spheroidal modes have angle-dependent radial deviation which create ripples in the Newtonian gravitational field that AGIS-LEO measures during its orbit.

In order to determine the AGIS-LEO response to free earth oscillations, we compute the perturbing potential $\phi$ resulting from an arbitrary surface displacement, which gives
\be \phi = -G \int_V \frac{\rho(\v{x}')\v{s}(\v{x}')\cdot(\v{x}-\v{x}')}{\left\|\v{x}-\v{x}'\right\|^3}dV' \label{Eqn: PhiFromDisplacement}\ee
\noindent where $\rho$ is the density profile of the Earth and $\v{s}(\v{x}') = \v{s}_{n,l,m}(r, \theta, \lambda, t)$ is the displacement associated with the $(n,l,m)^\text{th}$ Earth eigenmode \cite{DahlenGlobalSeismology}. Therefore the gravitational potential $\phi$ only depends on $(r, \theta, \lambda, t)$ and the initial amplitude and phase of the displacement.
\be \phi_{n,l,m}(r, \theta, \lambda, t) = \frac{R_E^{l+1}}{r^{l+1}} e^{i \omega_{n,l} (t + t_0)} Y_{l,m}(\theta,\lambda) f(R_E) \ee
\noindent where $f(R_E)$ is the integrated scale factor, $t_0$ is an arbitrary time offset, $\omega_{n,l}$ is the eigenfrequency, and $Y_{l,m}(\theta,\lambda)$ is shorthand notation for the fully normalized spherical harmonics that were previously defined, with $\bar{c}[m,n]$ and $\bar{s}[m,n]$ set to unity. The normal mode eigenfrequencies strongly depend on the Earth composition and are not well approximated by a simple analytic model. Therefore we use the MINEOS geophysics software package to calculate eigenfrequencies using the Preliminary Reference Earth Model (PREM) for $\rho(\v{x})$ \cite{MINEOS}.

We compute the phase shift for this time-dependent $\phi$ using perturbation theory, as described in the previous sections. We plot the response with diamond markers (green) on Fig. \ref{Fig:egm2008SphericalHarmonicResponse} where we have taken $f(R_E)$ such that the peak displacement for $s_{0,2,0}$ is $1~\text{cm}$, which was the minimum displacement of any location on Earth immediately after the 2004 Sumatra-Andaman earthquake \cite{JeffreyPark05202005}. This provides an upper bound to the $1~\text{year}$ integrated coherent signal response.

\subsubsection{Newtonian Gravity Spectral Signatures}
\begin{figure}
\includegraphics[width=7.0 in]{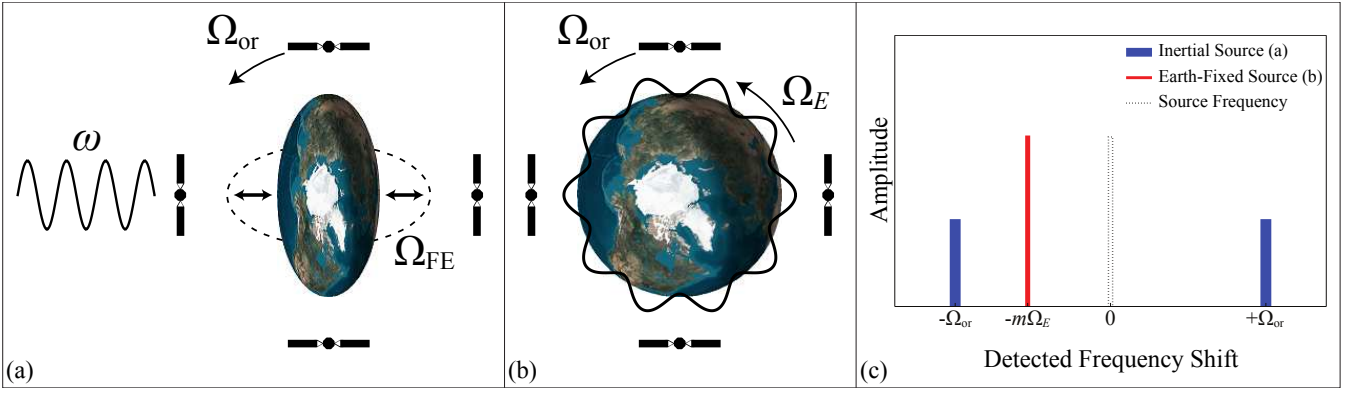}
\caption{ \label{Fig:GGfrequencies} Discrimination of Newtonian gravity gradient signals by their spectral signatures for equatorial satellite orbits.  (a) Inertial signal sources include gravity waves (at frequency $\omega$) and free-earth oscillations ($\Omega_\text{FE}$, ``football mode'' shown).  (b) Non-inertial Earth-fixed sources include the non-uniform geoid and terrain features, which rotate with the planet at a frequency of $\Omega_E$ (the spherical harmonic component pictured has $m=10$). (c) Amplitude spectral signatures of the effects pictured in (a) and (b).  For an equatorial orbit, inertial sources (blue, thicker bars) are amplitude modulated by the satellite's orbital frequency, $\Omega_\text{or}$, and appear as doublets at $\pm \Omega_\text{or}$ relative to the source frequency (gravity wave or free-earth oscillation).  For satellites in equatorial orbits, Earth-fixed sources would appear as signals at $m\Omega_\text{or}$ were it not for the rotation of the Earth.  With the Earth rotating at $\Omega_E$, these sources are downshifted by $m\Omega_E$ (red, thinner bar).  This picture is strictly correct only for an equatorial orbit, since a finite inclination angle would lead to doublet signatures even for Earth-fixed sources.}
\end{figure}

It is important to note that AGIS-LEO GW sources will persist over many satellite orbital periods, and can be discriminated from possible signals of terrestrial origin by their spectral signature (Fig. \ref{Fig:GGfrequencies}). Signal sources which are fixed to the Earth (such as those arising from the non-uniform geoid characterized by EGM2008) will be shifted in frequency by the Earth rotation rate if the satellite is in an equatorial orbit. However, signal sources which are in the inertial frame and not fixed to the rotating Earth, such as gravitational waves, will have their signal split into a doublet spaced by twice the orbital frequency. Free Earth oscillations are decoupled from the Earth rotation rate and also form doublets which are spectrally indistinguishable from gravity waves. However, free Earth oscillations are independently monitored by ground-based seismic networks and can be subtracted during signal post-processing.

\subsection{Laser Pointing Angle Jitter}
\label{subsec:PointingJitter}

In addition to the shot to shot laser pointing rotation rate error quantified by $\delta \Omega$, the pointing angle may vary from pulse to pulse. To determine the spectral response of the atoms to this jitter, we turn to an analysis procedure similar to that in Sec. \ref{SubSubSec:WavefrontCalculation}. In the particular case of laser pointing, the jitter is modeled as an oscillatory rotation angle $\delta \theta (t) = \overline{\delta \theta}(\omega) e^{-\imagI \omega t}$ that changes the direction of the wavevector $\v{k_{\text{eff}}}$ at each pulse time. Here, $\overline{\delta \theta}(\omega)$ is the angle amplitude spectral density and $\omega$ is the frequency of the rotation jitter.

\begin{figure}
\begin{center}
\includegraphics[width=3.0 in]{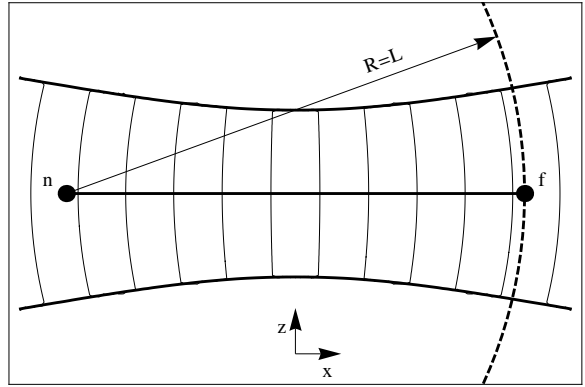}
\caption{ \label{Fig:gaussianCurvature} Pointing jitter insensitivity due to wavefront curvature of the Gaussian interferometer beams.  The two satellites are depicted as points labeled ``n'' and ``f'' for near and far, respectively.  The satellites are separated by a baseline $L$, and the atom interferometers are assumed to operate in close proximity to each of the satellites.  The interferometer beam waist is located at the midpoint between the two satellites and the beam has a Rayleigh range $x_R=L/2$ so that the radius of curvature of the wavefront at the far satellite is $R=L$.  As a result, the laser phase at the location of the far satellite is insensitive to pointing jitter caused by uncontrolled rotation of the beam about the near satellite.}
\end{center}
\end{figure}

A realistic model of laser pointing jitter must account for the finite curvature of the laser wavefronts at the position of the atoms.  To reduce sensitivity to pointing jitter, the curvature of the wavefronts may be chosen so that the phase of the beam at the position of the atom is minimally sensitive to beam rotation.  For a Gaussian beam propagating along the $x$-direction, the phase imprinted during one of the light pulses on an atom at position $(x,y,z)$ is given by
\be \Phi(x, y, z)=kx+\tfrac{k}{2 R(x)}(y^2+z^2)-\arctan{(x/x_R)} \ee
where $R(x)\equiv x + x_R^2/x$ is the radius of curvature of the wavefronts and $x_R$ is the Rayleigh range of the beam.  Notice that for an atom at position $x=x_R$, the radius of curvature is $R(x_R)=2x_R$.  Therefore, a beam with $x_R=L/2$ and a beam waist placed at the midpoint between the two atom interferometers separated by baseline $L$ will have a radius of curvature equal to the baseline distance at the position of the atoms (this is the same configuration as a confocal cavity).  This configuration results in a dramatic suppression of laser pointing sensitivity because the wavefront curvature at the position of the atom is locally the same as the curvature of a spherical wave originating from the point of rotation of the beam (see Fig. \ref{Fig:gaussianCurvature}).

To calculate the residual sensitivity to pointing jitter, we assume the laser configuration shown in Fig. \ref{Fig:gaussianCurvature} in which the beam waist is placed at the midpoint between the satellites and the Rayleigh range is $x_R=L/2$.  We consider the phase shift resulting from a two-photon transition (Raman or Bragg), and we assume that each of the two satellites serves as the point of origin for one of the two (counter-propagating) frequencies involved in the transition. For simplicity, we further assume that only one of the satellites experiences pointing jitter, so that only one of the counter-propagating Gaussian beams will have a variable propagation direction.  For the case that the near satellite is jittering (point ``n'' in Fig. \ref{Fig:gaussianCurvature}), the phase of the two-photon phase shift at the location of the far satellite (point ``f'' in Fig. \ref{Fig:gaussianCurvature}) is approximately given by
\be \Delta\Phi_\text{f}(x_\text{f}, z_\text{f})=2 k x_\text{f} + k z_\text{f}\left(\frac{x_\text{f}}{L}\right)\delta\theta(t) + k L\cdot O\!\left[\left(x_\text{f}/L\right)^2,\left(z_\text{f}/L\right)^2,(k L)^{-1}\right] \label{eqn:RotatedGaussianFar}\ee
where $x_\text{f}$ is the longitudinal position offset of the far atom away from the ideal Rayleigh range position, $z_\text{f}$ is the transverse offset, and we take $y=0$.  Notice that at the exact position of the Rayleigh range ($x_\text{f}=0$) the phase is (to first-order) independent of $\delta\theta(t)$.  Likewise, the two-photon phase shift at the location of the near satellite is approximately
\be \Delta\Phi_\text{n}(x_\text{n}, z_\text{n})=2 k x_\text{n} + k z_\text{n}\left(1 + \frac{x_\text{n}}{L}\right)\delta\theta(t) + k L\cdot O\!\left[\left(x_\text{n}/L\right)^2,\left(z_\text{n}/L\right)^2,(k L)^{-1}\right] \label{eqn:RotatedGaussianNear}\ee
where in this case $x_\text{n}$ is the longitudinal position of the near atom with respect to the point of rotation of the beam and $z_\text{n}$ is the transverse offset.  Note that the phase $\Delta\Phi_\text{n}$ at the position of the near atom is still sensitive to $\delta\theta(t)$ since the wavefront curvature cannot be matched at that distance from the rotation point.

\begin{table}
\setlength{\tabcolsep}{5pt}
\begin{tabular}{lcc}
& Differential Phase Noise & Maximum Term Size [$\mu\text{rad}/\sqrt{\text{Hz}}$]\\
\hline
\hline
$
 1  $ & $ \frac{27\sqrt{2}}{4} k_{\text{eff}} \delta x_n \frac{L}{R} \Omega_\text{or}^2 T^2 \chi(\omega T)  \overline{\delta\theta} $ & $ 108 $ \\ $
 2  $ & $ 9         k_{\text{eff}}^3\frac{\hbar^2}{L m^2}  \Omega_\text{or}   T^3  \sin(2\,\omega T) \,               \overline{\delta\theta}$ & $ 50 $ \\ $
 3  $ & $ 4  \,     k_{\text{eff}} \delta z_n                (7+8\cos(\omega T)) \sin^4\!\left(\frac{\omega T}{2}\right) \overline{\delta \theta} $ & $ 13 $ \\ $
 4  $ & $ 4  \,     k_{\text{eff}} (\delta x_\text{f}/L)\delta z_\text{f} \, (7+8\cos(\omega T)) \sin^4\!\left(\frac{\omega T}{2}\right) \overline{\delta \theta} $ & $ 4.2
$
\end{tabular}
\caption{Differential phase noise due to laser pointing jitter for AGIS-LEO.  The results are based on the triple-loop interferometer in Fig. \ref{Fig:FivePulseSequence}, and the phase difference assumes the gradiometer configuration shown in Fig. \ref{Fig:satelliteSchematic} between two atom interferometers separated by $L=30~\text{km}$.  The orbital radius $R$ and rotational rate $\Omega_\text{or}$ are for a $1000~\text{km}$ orbital altitude. The results are for the leader-follower orbit configuration.  Interrogation time is $T=4~\text{s}$ and the LMT beamsplitters have $\hbar k_{\text{eff}}=200\hbar k$. The frequency response of term 1 is given by $\chi(\omega T)\equiv\sqrt{155+100\cos(\omega T)- 39\cos(2\,\omega T)-16 \cos(3\,\omega T)} \sin ^2\left(\frac{\omega T}{2}\right)$.  The maximum term size is taken as the high-frequency limit of the enveloped response curves with the following parameters: $\overline{\delta \theta}= 1~\text{nrad}/\sqrt{\text{Hz}}$, $\delta z_n = 1~\mu\text{m}$, $\delta z_\text{f} = 1~\text{mm}$, $\delta x_\text{f}=10~\text{m}$, and $\delta x_n=10~\text{m}$.}
\label{tab:JitterTermList}
\end{table}

We perform the phase shift calculation outlined in Sec. \ref{subsection:ErrorModel} using the rotated Gaussian beam phase shifts of Eqs. (\ref{eqn:RotatedGaussianFar}) and (\ref{eqn:RotatedGaussianNear}).  This analysis assumes the leader-follower orbit configuration and the coordinate system shown in Fig. \ref{Fig:satelliteSchematic}.  The four largest terms of the RMS differential angle jitter response appear in Table \ref{tab:JitterTermList}.  The results have been simplified by enforcing the orbital conditions $g=\Omega_\text{or}^2 R$ and $T_\text{xx}=T_\text{yy}=-\tfrac{1}{2}T_\text{zz}=\Omega_\text{or}^2$.  Terms 3 and 4 are a consequence of the position dependence of $\Delta\Phi_\text{n}$ and $\Delta\Phi_\text{f}$.  Term 1 appears as the dominant response for the same analysis performed with plane waves instead of Gaussian beams; here the term is suppressed by $\delta x_n/L\sim 3\times 10^{-4}$ compared to the plane wave case as a result of the curved wavefront at the far satellite.  Finally, term 2 consists of equal contributions from both the near and the far satellite and is a result of the finite longitudinal extent of each atom interferometer coupled with transverse Coriolis deflections.

\begin{figure}
\begin{center}
\includegraphics[width=4.0 in]{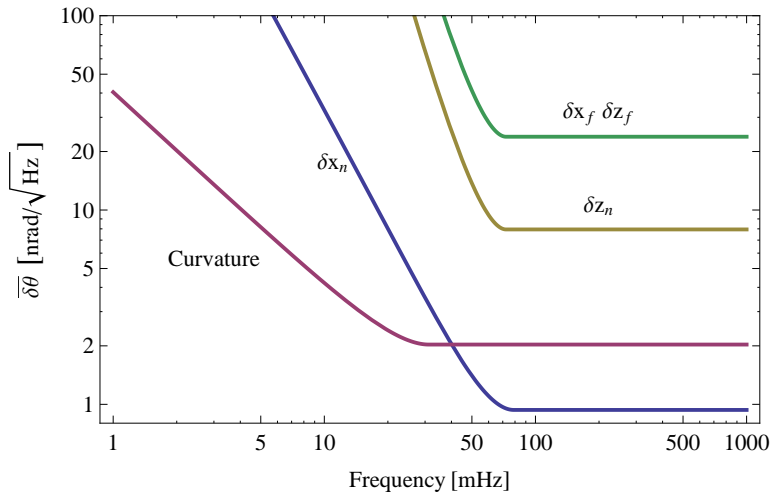}
\caption{ \label{Fig:pointingJitter} Interferometer laser pointing stability requirements.  Each curve corresponds to a different term from Table \ref{tab:JitterTermList}, labeled by their corresponding position offsets.  ``Curvature'' refers to Term 2.  All parameters are the same as in Table \ref{tab:JitterTermList}.}
\end{center}
\end{figure}

The angle jitter requirements implied by each of these terms is shown in Fig. \ref{Fig:pointingJitter} as a function of jitter frequency.  In the relevant gravitational wave bandwidth, the pointing error must be $<1~\text{nrad}/\sqrt{\text{Hz}}$ in order for the phase noise contribution to be less than atom shot noise of $10^{-4}~\text{rad}/\sqrt{\text{Hz}}$. Table \ref{tab:SatelliteControl} summarizes the satellite control constraints.

In this analysis we have assumed that the photon recoil effects due to Gaussian curvature are negligible. The wavevector $\vec{k}$ is the gradient of the local phase during the atom-light interaction. The finite beam curvature causes the effective wavevector $\vec{k}$ to point slightly away from the interferometer axis, at off-axis interaction points. This changes the direction of the momentum kick imparted to the atom, which leads to additional transverse motion. These trajectory modifications are small compared to other deflections present in the problem, such as those caused by Coriolis forces. However, the effect of these wavevector perturbations can be computed and this analysis is currently ongoing.

The results of the pointing jitter analysis are mostly unchanged for the inclined-great-circles orbit configuration.  The main difference is that the transverse Coriolis deflections are greatly suppressed for the inclined-great-circles.  As a result, term 2 in Table \ref{tab:JitterTermList} no longer appears in the results.  However, the jitter requirement is still $< 1 \text{ nrad/}\sqrt{\text{Hz}}$ in the GW band for this orbit configuration.

\section{Short-Baseline Configurations}
\label{ShortBaseline}

A proof-of-principle AGIS instrument could be space tested on the International Space Station (AGIS-ISS) or on an independent satellite employing booms to house the interferometer regions.  While such an instrument would be limited to significantly shorter baselines, thereby decreasing sensitivity to the scientifically interesting gravitational waves mentioned previously, it would provide useful technology development for a full AGIS-LEO mission.

In addition to serving as a testbed for the light-pulse atom interferometry sequences that will be used for AGIS-LEO, a short-baseline instrument could be used to implement lattice-hold atom interferometry, a recently-proposed alternative to light-pulse atom interferometry \cite{ref:kovachy}. Lattice-hold atom interferometry involves the use of optical lattices to continuously control the trajectories of the two arms of an atom interferometer.  To perform optimally, optical-lattice manipulations of the atoms require a larger two-photon Rabi frequency (which is proportional to laser intensity) than may be available in the AGIS-LEO configuration.  This is because available laser power is finite, and so long baselines, which require larger beams to mitigate diffraction, have limited intensity. With the smaller beam waist allowed by a shorter baseline, the desired intensities for lattice-hold interferometry are more easily attainable.  The potential improvements in sensitivity of this promising new approach to atom interferometry suggest that it might even be capable of observing gravitational waves with ISS-scale baselines.  Regardless, it requires the same hardware as light-pulse atom interferometry, with only software modifications needed to change the durations and frequencies of the applied laser light.  Thus, a short-baseline mission could provide a space demonstration of both light-pulse atom interferometry and lattice-hold atom interferometry, with the possibility of direct gravitational wave detection.

In what follows, we provide a brief conceptual overview of lattice-hold interferometry as it could be applied to an ISS or other short-baseline satellite mission.  Much of the analysis we have performed for a light-pulse interferometer geometry applies also to the case of a confined lattice-hold interferometer.  In this section, we focus on the aspects of the analysis that must be modified for confined interferometer geometries and short baselines.

\subsection{Lattice-Hold Interferometer Gravitational Wave Phase Shift}
\label{sec:latHoldPhase}

\begin{figure}
\begin{center}
\includegraphics[width=2 in]{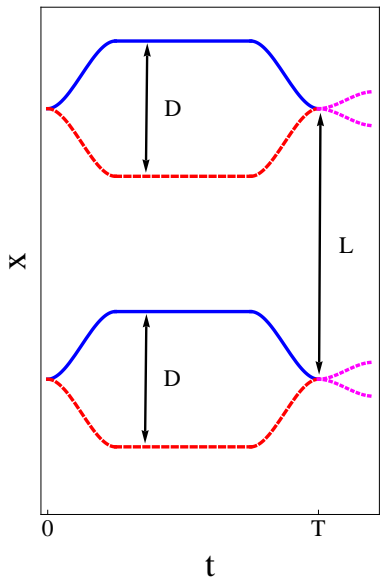}
\caption{ \label{Fig:LatticeHold} Lattice-hold interferometers in a differential configuration that could be realized on the ISS.  The baseline distance between the two interferometers is $L$, the wavepacket separation in each interferometer is $D$, and the interrogation time is $T$.  The gravitational-wave-induced phase difference between the interferometers is given by Eq. (\ref{Eqn: GW phase lattice trig}).}
\end{center}
\end{figure}

In comparison to light-pulse atom interferometry, lattice-hold atom interferometry offers potentially larger gravitational wave sensitivities for short baseline configurations.  The gravitational-wave-induced phase difference between two lattice-hold interferometers is:
\be
\Delta\phi_\text{GW} \sim \frac{m}{\hbar} \omega L  D h \sin \left (\omega T/2 \right) \sin{\theta_\text{GW}}
\label{Eqn: GW phase lattice trig}
\ee
\noindent where $m$ is the mass of the atom, $h$ is the gravitational wave strain, $\omega$ is the gravitational wave frequency, $T$ is the interrogation time, $L$ is the baseline distance between the two interferometers, and $D$ is the wavepacket separation in each interferometer.  The phase $\theta_\text{GW}$ of the gravitational wave is defined as in Sec. \ref{Sec:gwphase}.  The interferometer configuration is illustrated in Fig. \ref{Fig:LatticeHold}.  Equation (\ref{Eqn: GW phase lattice trig}) follows from a fully relativistic calculation in which the space-time trajectories along which the lattices guide the atoms are determined by solving for the intersection points of the null geodesics traveled by the photons forming the lattices.  These intersection points are calculated using the general method presented in \cite{GRAtom}.  An interesting feature of the phase difference in Eq. (\ref{Eqn: GW phase lattice trig}) is that if we set $D \sim L$, it scales quadratically with $L$, suggesting the possibility of scientifically interesting sensitivities with only modest baselines.

\subsection{Lattice-Hold Interferometer Gravitational Wave Sensitivity}

To illustrate the promising sensitivity of lattice-hold interferometer gravitational wave detectors over short baselines, we consider an example set of feasible parameters.  For a baseline of $L \sim 10 \; \text{m}$, a wavepacket separation of $D \sim 10 \; \text{m}$, a gravitational wave frequency of $\omega \sim 2 \pi \times 1 \; \text{Hz}$, and a phase sensitivity of $\sim 10^{-4} \; \text{rad}  / \sqrt{\text{Hz}}$, a lattice-hold interferometer configuration using Rb atoms has a strain sensitivity of $h\sim 10^{-16}/\sqrt{\text{Hz}}$.  For laser intensities of $\sim 100 \;\text{W}/\text{cm}^2$, a sufficiently large two-photon Rabi frequency ($\sim 2 \pi \times 60 \; \text{kHz}$) can be achieved to efficiently pull apart and recombine the atomic wavepackets over the $\sim 10 \; \text{m}$ baseline in $\sim 1 \; \text{s}$, while the spontaneous emission rate can be kept sufficiently low ($\sim 0.1 \; \text{s}^{-1}$) to avoid significant losses.

The parameters described allow for a scientifically interesting gravitational wave strain sensitivity for an apparatus that could be fit on the ISS.  Lattice-hold schemes are also promising for ground-based detectors.  For instance, such a detector could be realized in the $10 \; \text{m}$ atomic fountain currently being constructed at Stanford \cite{GRAtom}.

\subsection{Environmental Considerations near the International Space Station}

As discussed previously, most of the environmental factors of Sec. \ref{sec:EnvCons} become more problematic at lower altitudes.  At $350~\text{km}$ above the surface of the Earth (the approximate orbital altitude of the ISS), the number density of atomic oxygen is around $10^8~\frac{\text{atoms}}{\text{cm}^3}$, at least two orders of magnitude worse than required \cite{MSISE90}.  Thus, the AGIS-ISS instrument would need to have an enclosed vacuum tube to protect the interferometry region, though the exterior vacuum environment allows the tube walls to be thin and non-structural.  This tube would also act as a sunshield, increasing uptime by eliminating the need to operate only in the Earth's shadow.  It could even provide the base for a magnetic shield if one were necessitated by the lower orbit or the proximity of the ISS.  Further study is required to determine the impact of gravity gradients and Coriolis effects in the lower ISS orbit and to develop a full systematic error budget for AGIS-ISS, although preliminary calculations indicate that the transverse confinement provided by red-detuned laser beams for a lattice-hold configuration can significantly suppress the sensitivity of the instrument to Coriolis effects.  AC Stark shifts will be an important factor in choosing the optimal beam waist and in specifying the necessary level of intensity stabilization.  These shifts can cause non-common noise in the interferometer phase that depends both on the degree of divergence of the laser beams over the interferometer region and on the size of laser intensity fluctuations.  Finally, to mitigate vibrational noise from the ISS itself, one option is to tether the instrument to the space station, rather than to attach it directly.

\subsection{Three-Axis Boom-Based Configuration}
\label{Boom}

\begin{figure}
\begin{center}
\includegraphics[width=3in]{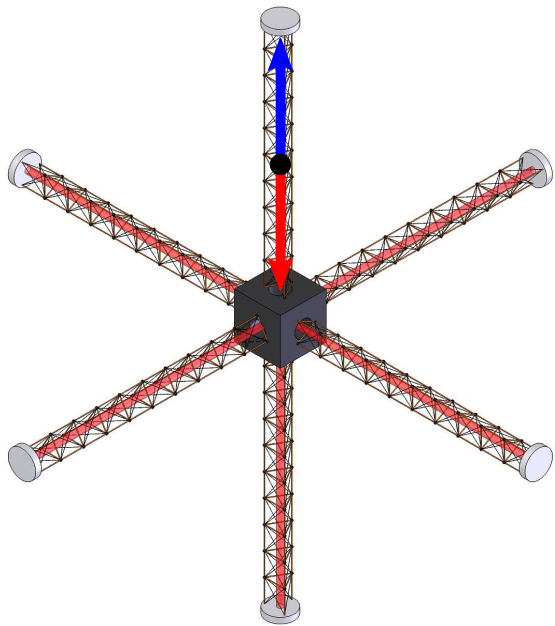}
\caption{\label{Fig:cubeWithBooms} Boom configuration for a three-axis gravitational wave detector, with retroreflectors mounted at the end of each boom.  The shaded red lines indicate the optical path, and the arrows on the upper boom indicate an atom interferometer fully filling its boom (specifically the trajectories of the atoms during the first half of the first loop of a sequence, as in Fig. \ref{Fig:FivePulseSequence} or Fig. \ref{Fig:LatticeHold}).  The advantages of a boom configuration are discussed in Sec. \ref{Boom}.  Note that the figure is not to scale, that the vacuum bellows in each boom are hidden for clarity, and that only one of the six interferometer regions is indicated with arrows.}
\end{center}
\end{figure}

A promising configuration for short-baseline detectors involves the use of six booms originating from a central module to form interferometer pairs along three orthogonal axes, with each boom housing a single interferometer (see Fig. \ref{Fig:cubeWithBooms}).  The laser beams can be aligned in a racetrack configuration through the use of retroreflecting elements (e.g., corner cubes) at the end of each boom and pentaprisms inside the central module to allow for 90 degree turns, enabling us to simultaneously address all six interferometers with a common set of frequency ramps.  The central module would house the atom source and optics, with bellows extending inside each boom to establish vacuum.  As mentioned in the previous section, a magnetic shield can enclose each of the interferometer regions or be incorporated directly into the vacuum bellows.  Compact, self-deploying booms are available from commercial manufacturers \cite{ABLE}.

A key advantage of a boom configuration is that the three-axis measurement allows for the mitigation of background signals due to fluctuations in centrifugal force gradients, which arise from rotational noise.  For a single-axis sensor, such fluctuations have the same signature as a gravitational wave.  However, correlation measurements between three orthogonal sensors would allow for these fluctuations to be distinguished from a gravitational wave.

A boom configuration can either operate as a free flyer or be attached to the ISS, with the capability to implement both light-pulse and lattice-hold interferometry due to their identical hardware requirements.  In either case, it is desirable for each interferometer to fill its respective boom (i.e., the wavepacket separation should be nearly the length of a boom), so that the maximal gravitational wave sensitivity allowed by the instrument is achieved.   For a free flyer configuration, it may be possible to use boom lengths of up to $\sim 100$ m.  With such a boom length and with $\hbar k_\text{eff}=200 \hbar k$ beam splitters, the interferometers can fill the booms with interrogation times of $\sim 100$ s.

For a free flyer, one axis could be made effectively inertial by orienting it perpendicular to the orbital plane, suppressing effects arising from rotation bias and Coriolis deflections as is done for the IGC configuration (see Sec. \ref{Sec:SatelliteConfiguration}).  With the use of thrusters, it may be possible to make all three axes effectively inertial.

An additional scientific motivation for the boom configuration is that the three-axis measurement can aid in the performance of certain tests of general relativity.  For example, the boom configuration appears to be an ideal platform for measuring the nonzero divergence of the gravitational field in free space predicted by general relativity \cite{GRAtom}.

\section{Secondary Objectives}

Although variations in Earth's gravity gradients are a noise source for the GW detector, pulse sequences designed to measure these variations can provide important geophysical information including data regarding the flow of water on the Earth that is important for understanding weather, the climate, and oceans. Similar arguments have provided motivation for satellite missions such as the CHAllenging Mini-satellite Payload (CHAMP), the Gravity Recovery and Climate Experiment (GRACE), and the Constellation Observing System for Meteorology, Ionosphere and Climate (COSMIC).  The method works: GRACE's monthly gravitational maps have improved our understanding of ocean circulation \cite{GRACE_Oceans}, polar ice loss \cite{GRACE_IceLoss}, and water flow of large basins like the Amazon \cite{GRACE_WaterBasin}.

Furthermore, if the atom clouds are prepared in a magnetically-sensitive state, $\left|m \neq 0\right>$, the gravity wave detector becomes a detector of the local magnetic field gradient at each of the interferometers. Similar to the gravity gradiometer, the transformed detector would produce a map of the temporal variation in the Earth's magnetic gradients. Data collected by previous satellite missions--\O rsted, CHAMP, and Satellite Argentina Cientificas - C (SAC-C)--have been used to model the static and time-varying geomagnetic field \cite{OCS_Geomag} and mantle conductivity \cite{OCS_Mantle}.

The modifications necessary to make these measurements do not require different hardware, merely different software. With this unique flexibility of atom interferometric measurements, AGIS-LEO can provide complementary maps of temporal variations in the Earth's gravity gradients and magnetic field gradients.

\section{Summary}

Atom interferometry offers a means to sensitively measure gravitational waves with short baselines and in low Earth orbit. With a strain sensitivity of $< 10^{-18}/\sqrt{\text{Hz}}$ in the $50~\text{mHz}$ - $10~\text{Hz}$ frequency band, the AGIS-LEO instrument can detect gravitational wave signals inaccessible to ground-based alternatives. Such information could impact our understanding of the universe from the mergers of nearby white dwarf binaries to phase transitions and reheating at the earliest of times. Extensive theoretical analysis of noise sources performed for solar orbits, \cite{AGIS}, has been adapted to the new conditions of low Earth orbit. We have analyzed possible environmental noise sources from near-Earth magnetic fields and gravity gradients and have found them to fall below the necessary threshold. Laser requirements--phase noise, frequency stability, and wavefront--are all achievable with current technology.  Rotation effects, which are stronger in low Earth orbits due to the short orbital period, can be mitigated through the invaluable flexibility of atom interferometry: the choice of interferometer pulse sequence.

An apparent strength of the AI approach is that many of the instrument and system level requirements can be validated on the ground, in suitably configured terrestrial laboratories.  For example, the Stanford 10 m free fall tower will enable characterization of LMT optical sequences, wavefront and laser jitter requirements, magnetic field sensitivity, vacuum sensitivity, laser cooling protocols, kinematic requirements for the atom sources, Coriolis and gravity gradient sensitivities, etc.

Finally, atom interferometry has the potential for even higher precision. With longer baselines and larger beamsplitters, the sensitivity of pulsed atom interferometers to gravitational waves can be increased by multiple orders of magnitude. With shorter baselines, lattice-hold atom interferometers, now in development, have the capability to provide high sensitivity.  In fact, an ISS-scale test mission for AGIS using lattice-hold atom interferometry might be capable of directly detecting gravitational waves, while simultaneously flight testing the technology for a full AGIS-LEO science mission.

In short, AGIS is a sensitive and scalable instrument that can open the door to information about our universe that is currently encoded in gravitational waves.

\begin{acknowledgments}

We thank Peter Bender for valuable discussions.  TK acknowledges support from the Fannie and John Hertz Foundation. TK, SD and AS acknowledge support from the NSF GRFP.  AS acknowledges support from an Anne T. and Robert M. Bass Stanford Graduate Fellowship and a DoD NDSEG Fellowship.  SR is supported by the DoE Office of Nuclear Physics under grant DE-FG02-94ER40818. SR is also supported by NSF grant PHY-0600465.  PB acknowledges support from a Fulbright Fellowship and the iXCore Foundation.

\end{acknowledgments}

\bibliography{agisLeo}

\end{document}